\documentclass[letterpaper,twocolumn,10pt]{article}
\usepackage{usenix2019_v3}

\usepackage{tikz}
\usepackage{amsmath}
\usepackage{filecontents}
\usepackage{comment}
\usepackage{url}
\usepackage{multirow}
\usepackage{amssymb}
\usepackage{pifont}
\usepackage{xparse}
\usepackage{xspace}
\usepackage{listings}
\usepackage{soul}
\usepackage{flushend}
\usepackage{balance}
\usepackage{nopageno}

\usepackage[toc,page]{appendix}
\newcommand{\name}{Hermes Attack\xspace}
\newcommand{\mypara}[1]{\vspace{2pt}\noindent\textbf{{#1. }}}

\usepackage{tikz}
\usetikzlibrary{calc}

\newcommand{\circled}[1]{
    \setbox0=\hbox{#1}%
    \dimen0\wd0%
    \divide\dimen0 by 2%
    \begin{tikzpicture}[baseline=(a.base)]%
        \useasboundingbox (-\the\dimen0,0pt) rectangle (\the\dimen0,1pt);
        \node[circle,draw,outer sep=0pt,inner sep=0.05ex, minimum size=11pt] (a) {\small{#1}};
    \end{tikzpicture}
}

\begin{document}
%-------------------------------------------------------------------------------

\date{}
%\title{\Large \bf Hermes Attack: Steal DNN Models with Zero Inference Accuracy Reduction}
\title{\Large \bf Hermes Attack: Steal DNN Models with {Lossless Inference Accuracy}}
\author{{\rm Yuankun Zhu\thanks{This work was mainly done during the internship at Baidu.}}\\ The University of Texas at Dallas\\{\rm yuankun.zhu@utdallas.edu} \and {\rm Yueqiang Cheng*}\\ Baidu Security\\{\rm chengyueqiang@baidu.com} \and {\rm Husheng Zhou}\\ VMware \\{\rm zhusheng@vmware.com} \and {\rm Yantao Lu}\\ Syracuse University\\{\rm ylu25@syr.edu}}
\maketitle

%-------------------------------------------------------------------------------
\begin{abstract}
%-------------------------------------------------------------------------------
Deep Neural Network (DNN) models become one of the most valuable enterprise assets due to their critical roles in all aspects of applications.
With the trend of privatization deployment of DNN models, the data leakage of the DNN models is becoming increasingly severe and widespread. 
All existing model-extraction attacks can only leak parts of targeted DNN models with low accuracy or high overhead. 
In this paper, we first identify a new attack surface -- unencrypted PCIe traffic, to leak DNN models.
Based on this new attack surface, we propose a novel model-extraction attack, namely \emph{\name}\footnote{Hermes is the master of thieves and the god of stealth~\cite{hermes}.}, which is the first attack to fully steal the \emph{whole} victim DNN model.
The stolen DNN models have the same hyper-parameters, parameters, and semantically identical architecture as the original ones. 
It is challenging due to the closed-source CUDA runtime, driver, and GPU internals, as well as the undocumented data structures and the loss of some critical semantics in the PCIe traffic. Additionally, there are millions of PCIe packets with numerous noises and chaos orders. 
Our \name~addresses these issues by massive reverse engineering efforts and reliable semantic reconstruction, as well as skillful packet selection and order correction.
We implement a prototype of the \name, and evaluate two sequential DNN models (i.e., \texttt{MINIST} and \texttt{VGG}) and one non-sequential DNN model (i.e., \texttt{ResNet}) on three NVIDIA GPU platforms, i.e., \texttt{NVIDIA Geforce GT 730}, \texttt{NVIDIA Geforce GTX 1080 Ti}, and \texttt{NVIDIA Geforce RTX 2080 Ti}.
The evaluation results indicate that our scheme can efficiently and completely reconstruct ALL of them by making inferences on any one image.
Evaluated with \emph{Cifar10 test dataset} that contains $10,000$ images, the experiment results show that the stolen models have the same inference accuracy as the original ones (i.e., {lossless inference accuracy}). %(i.e., \emph{zero} accuracy reduction).
%We will open-source these reverse engineering results, hoping to benefit the entire community.
%trending: privatization deployment
%citation adversary attack
\end{abstract}

%-------------------------------------------------------------------------------
\section{Introduction}
\label{sec:intro}
%-------------------------------------------------------------------------------
%big picture, why AI is important,
%narrow down scope, AI deployment on PC, software protect
%related work, limitation
%our attack overview
%motivation,intro,result,platform
%Contribution 3~4
%Structure

Nowadays, Deep Neural Networks (DNNs) have been widely applied in numerous applications from various aspects, such as Computer Vision~\cite{xie2012image,cirecsan2012multi}, Speech Recognization~\cite{graves2013speech,hinton2012deep}, Natural Language Processing~\cite{collobert2008unified}, and Autonomous Driving, such as Autoware~\cite{kato2015open}, Baidu Appolo~\cite{appolo}, Tesla Autopilot~\cite{tesla}, Waymo~\cite{Waymo}. These applications indicate the principle role of DNNs in both industry and academic areas. 
Compared to other machine learning technologies, DNN stands out for its human-competitive accuracy in cognitive computing tasks, and capabilities in prediction tasks~\cite{lecun2015deep,schmidhuber2015deep}.
%Compared to other machine learning technologies, the main reason why DNN stands out is its excellent achievement of prediction tasks~\cite{lecun2015deep,schmidhuber2015deep}.
The accuracy of a DNN model is highly dependent on internal architecture, hyperparameters, and parameters, which are typically trained from a TB datasets~\cite{deng2009imagenet,xiao2010sun} with high training costs. For instance, renting a v2 Tensor processing unit (TPU) in the cloud is \$4.5 per hour, and one full training process would cost \$400K or higher~\cite{devlin2018bert,raffel2019exploring}. 
Therefore, the importance of protecting DNN models is self-evident. 

Over the last few years, privatization deployments~\cite{JDAI, BaiduAI} are becoming a popular trending for giant AI providers. 
The AI providers have private high-quality DNN models, and would like to sell them to other companies, organizations and governments with a license fee, e.g., million dollars per year. 
This privatization-deployment situation further exacerbates the risk of model leakage.
There have been many DNN extraction works proposed in the literature~\cite{hu2019neural,yan2020cache,hua2018reverse,xiang2019open,wei2018know,duddu2018stealing,wang2018stealing,oh2019towards,tramer2016stealing,shokri2017membership}. 
All of them use either a search or prediction method to recover DNN models.
For the search based schemes~\cite{hua2018reverse,yan2020cache}, they can only obtain existing models but not customized models. Besides, the performance of their searching processes is particularly low.
The prediction based schemes~\cite{duddu2018stealing,hu2019neural,xiang2019open} result in a significant drop in inference accuracy. Most importantly, all of these attacks are \emph{not} able to reconstruct the whole DNN model. 
Thus, until now, most people still have the illusion that the model is safe enough or at least the leakage is limited and acceptable.

In this paper, we first observed that the attacker in the model privatization deployment has physical access to GPU devices, making the PCIe bus between the host machine and the GPU devices become a new attack surface. 
Even if the host system and the GPU are well protected individually (e.g, using Intel SGX protect DNN model on the host and never sharing GPU with others), the attacker still has the chance to snoop the unencrypted PCIe traffic to extract DNN models.
Based on this critical observation, we propose a novel black-box attack, named \name, to entirely steal the whole DNN model, including the architecture, hyper-parameters, and parameters.

%The idea behind our work comes from the heterogeneous CPU-GPU platform features: Most DNNs use GPUs to accelerate the both training and inference processes. CPU and GPU are connected via PCIe in heterogeneous platforms. If we can intercept the PCIe traffic, it's possible to extract some internal information of DNNs and further reconstruct the whole model.

It is challenging to fully reconstruct DNN models from PCIe traffic even if we can intercept and log all PCIe packets due to the following three aspects. First, the CUDA runtime, GPU driver, and GPU internals are all closed source, and the critical data structures are undocumented. The limited public information makes the reconstruction extremely difficult.
Second, some critical model information, such as the information about layer type, is lost in the PCIe traffic. Without this critical information, we cannot fully reconstruct the whole DNN model. 
At last, there are millions of PCIe packets with numerous noises and chaos orders. 
Based on our experiments, only 1\% to 2\% of all captured PCIe packets are useful for our model extraction work.

To address the above challenges, we design our \name into two phases: offline phase and online phase. The main purpose of the offline phase is to gain domain knowledge that is not publicly available. Specifically, we recover the critical data structures, e.g., GPU command headers, using a large number of reverse engineering efforts to address challenge 1.
We address challenge 2 based on a key observation: \emph{each layer has its own corresponding {unique} GPU kernel}.
Thus, we identify the mapping relationship between the kernel (binaries) and the layer type in the offline phase with known layer type and selected white-box  models. 
We put all these pair information into a database, which will benefit the runtime reconstruction.
In the online phase, we run the victim model and collect the PCIe packets. 
By leveraging the PCIe specification and the pre-collected knowledge in the database, we correct the packet orders, filter noises, and fully reconstruct the whole DNN model, to address challenge 3.

To demonstrate the practicality and the effectiveness of \name, we implement it on three real-world GPU platforms,  i.e., \texttt{NVIDIA Geforce GT 730}, \texttt{NVIDIA Geforce GTX 1080 Ti}, and \texttt{NVIDIA Geforce RTX 2080 Ti}.
The PCIe snooping device is Teledyne LeCroy Summit T3-16 PCIe Express Protocol Analyzer~\cite{pcieanalyzer}.
%We adapt a customized model as the victim DNN model. This model is implemented with Keras and Tensorflow, comprising eight layers and four activation functions. The parameters of this model are trained from MNIST dataset.
We choose two sequential DNN models - MNIST~\cite{lecun2010mnist} and VGG~\cite{simonyan2014very}, and one non-sequential model - ResNet~\cite{he2016deep}. These three pre-trained victim models are used for interference by Keras framework~\cite{keras} with Tensorflow\cite{abadi2016tensorflow} as the backbone.
The attack experiments indicate that \name is effective and efficient: (1) randomly given one image, we can completely reconstruct the whole victim model within 5 -- 17 minutes; and (2) the reconstructed models have the same hyper-parameters, parameters, and semantically identical architecture as the original ones.
In the inference accuracy experiments, we test each reconstructed model with $10,000$ images from public available test datasets~\cite{lecun2010mnist,krizhevsky2014cifar}. The results show that the reconstructed models have exactly the same accuracy as the original ones (i.e., {lossless inference accuracy}). %zero accuracy reduction).

\mypara{Contributions} In summary, we make the following contributions in this paper:
\begin{itemize}
\item We are the first to identify the PCIe bus as a new attack surface to steal DNN models in the model-privatization deployments, e.g., smart IoT, autonomous driving and surveillance devices.

\item We propose a novel \name, which is the first black-box attack to fully reconstruct the \emph{whole} DNN models. \emph{None} of the existing model extraction attacks can achieve this.

\item We disclose a large number of reverse engineering details in reconstructing architectures, hyper-parameters, and parameters, benefiting the whole community.

\item We have demonstrated the \name on three real-world GPU platforms with sequential and non-sequential models. The results indicate that the \name can handle MNIST, VGG and ResNet DNN models and the reconstructed models have the same inference accuracy as the original ones.
\end{itemize}

%The rest of this paper is organized as follows. Section \ref{background} describes the background of DNN, GPU, and PCIe. Section \ref{attack} presents both overview and the implementation detail of \name. Section \ref{evaluation} evaluates the effectiveness of our attack. Section \ref{discussion} discusses some applicable attack methods in the case of other DNN models, as well as defense approaches. Section \ref{relatedwork} records the related work, and Section \ref{conclusion} concludes the paper.

%-------------------------------------------------------------------------------
\section{Background} \label{background}
%-------------------------------------------------------------------------------

\subsection{DNN Background}

Deep Neural network (DNN) is a sub-area of machine learning in artificial intelligence that deals with algorithms inspired from the biological structure and functioning of a brain. 
DNN is used to model both linear and non-linear relationships between the input $x$ and the output $y$, learning to approximate an unknown function $f(x) = y$. 
A DNN model is represented as a hierarchical organization of connected layers with a certain level of complexity between the input data and resultant output.
DNNs are used in two phases, i.e., training and inference.
The training process is computationally heavy and needs a large amount of data. 
With a series of feed-forward matrix computations on given input data, the resultant output is computed through a loss function against ground truth. The weights of the network are updated accordingly based on error back-propagation. The training is done once passing through all of the training samples.
The inference is the phase in which a trained model is used to infer real-world data.
Terminologies used in the rest of this paper are described as follows.

\vspace{1mm}\noindent\textbf{Architecture:}
Neural network architecture consists of a number of layers, types/dimensions for each layer, and connection topology among layers. The connections between layers can be either sequential or non-sequential.
Sequential connection means layers are stacked and every layer take the only output of the previous layer as the input. Non-sequential connection denotes the model may include shortcuts, branches, or shared layers\cite{keras,yan2020cache}.

\vspace{1mm}\noindent\textbf{Hyper-parameters:}
Hyper-parameters are the parameters used to control the training process, which do not belong to the trained model and cannot be estimated from training data.
There are many hyper-parameters such as learning rate, regularization factors, momentum coefficients, number of epochs, batch size, etc.

\vspace{1mm}\noindent\textbf{Parameters:}
Parameters are configuration variables of the trained model, whose values are derived via training. Model parameters includes weights and bias in DNNs. Throughout the paper, when we mention ``parameters'', we mean DNN model parameters instead of ``arguments''.

\subsection{GPU Working Mechanism}
\begin{figure}[t]
\centering
\includegraphics[scale=0.8]{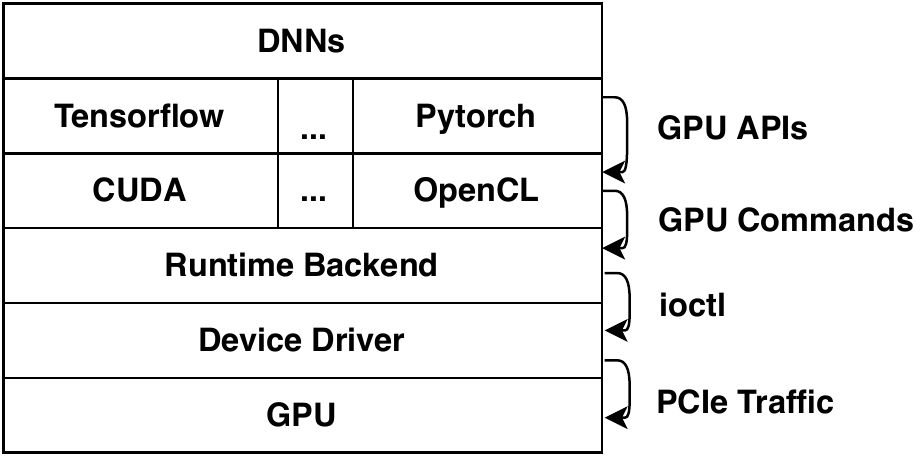}
\caption{\textbf{Typical DNN System Stack.} DNNs are usually implemented with deep learning frameworks, e.g., Tensorflow, Pytorch, and Caffe. These frameworks invoke the GPU runtime frontend like CUDA by calling APIs. The runtime frontend converts these APIs to GPU commands and sends them to the runtime backend, which then sends the received commands to the device driver through  $ioctl$. The device driver submits these commands to GPU hardware via PCIe.}
\label{fig:GPU}
\vspace{-1mm}
\end{figure}

%What is GPU:
%Since both training and inference are computationally intensive processes, they give GPUs the upper hand to speed up the process. On the one hand, training demands high throughput. DNNs are structured in a very uniform manner as thousands of neurons at each layer of the neural network perform the same computation. It is similar to how GPUs perform instruction computation. Besides, GPUs have the advantage of parallel computation capabilities and high bandwidth to retrieve from memory. Given the massive parallelism of GPUs, it is energy efficient to batch hundreds of training inputs and perform propagation simultaneously to amortize the cost if loading network weights from GPU memory across many inputs. On the other hand, the primary performance goal of inference is latency as automated services require to respond in real-time, inference batches a small number of inputs to minimize the end-to-end response time. Therefore, GPUs are specifically suitable for DNN applications. 

Adding sufficient DNN layers to guarantee high inference accuracy may easily explode the computation demand~\cite{cui2016geeps}. Currently, major DNN frameworks mainly rely on employing GPUs to satisfy the need, since GPUs enable orders of magnitude acceleration and more energy-efficient execution for many DNN related computations. According to their architecture, modern GPUs can be divided into integrated GPUs that lie on the same die of CPUs and discrete GPUs which are connected to CPU via PCIe. Integrated GPUs are more energy-efficient but less powerful, which is often seen in embedded systems and mobile devices. In this paper, we focus on discrete GPUs since they dominate the markets of AI and machine learning for their computation powers. Some terminologies used in this paper are described as follows.

%For discrete GPUs, GPU and CPU connected via PCIe, and both CPU and GPU have their own memory, which is usually more powerful and generally used in personal computers and workstations. For integrated GPUs, the GPU is integrated with CPU on the same chip and share a single memory, which is usually more energy-efficient and is used in embedded systems and mobile phones. In this paper, we focused on discrete GPU architectures. There are several keywords in our paper:

\vspace{1mm}\noindent\textbf{CUDA} is a parallel computing architecture provided by NVIDIA for GPUs\cite{nvidia2011nvidia}, which includes compilers, user space libraries, and kernel space drivers. Employing CUDA for a very simple GPU accelerated program usually involves three procedures: copying input data from main memory to GPU memory, launching computations on GPU, and transferring back the resultant output from GPU memory to main memory. 

\vspace{1mm}\noindent\textbf{Kernel} is a piece of code that is compiled into hardware-specific executable and runs on GPU hardware to do the actual computation. Throughout the paper, when we mention ``kernel'' we mean ``GPU kernel'' instead of OS kernel. 
In CUDA, kernels are compiled by nvcc compiler~\cite{nvcc} into CUDA Fatbin and embedded into a dedicated section of host executable file. During runtime, sets of GPU instructions are loaded onto GPU and launched when specific CUDA APIs are called (e.g., cudaLaunchKernel).

\vspace{1mm}\noindent\textbf{Commands} are encoded using distinct instruction sets with kernels, which are used to control data copy, kernel launch, initialization, synchronization, etc. In this paper, we use ``GPU command'' to indicate a set of GPU hardware instructions that complete an atomic CUDA operation.
Each GPU command consists of two parts: the header and the data. The header contains the type of this command and the data size. The data field comprises values passed to this command. We named the data movement command as \texttt{$D$ command} and the kernel launch command as \texttt{$K$ command} in the rest of the paper. 

% Both of them are based on a conception of “pushbuffer”: an area of memory that user fills with commands and tells % PFIFO to process. The pushbuffers are then assembled into a “command stream” consisting of 32-bit words that make
% up “commands”. In NV4-style DMA mode, the pushbuffer is always read linearly and converted directly to command
% stream, except when the “jump”, “return”, or “call” commands are encountered. In IB mode, the jump/call/return
% commands are disabled, and command stream is instead created with use of an “IB buffer”. The IB buffer is a circular
% buffer of (base,length) pairs describing areas of pushbuffer that will be stitched together to create the command stream.
% NV4- style mode is available on NV4:GF100, IB mode is available on G80+.

\vspace{1mm}\noindent\textbf{GPU Accelerated DNN Platform} is depicted as Figure~\ref{fig:GPU}, which includes DNN frameworks, user space libraries, kernel space drivers, and the hardware. High level computation tasks of DNN are finally converted to low level PCIe packets, which is the attack surface we are targeting in this paper.

\subsection{PCIe Protocol}

\begin{figure}[t]
\centering
\includegraphics[width=\linewidth]{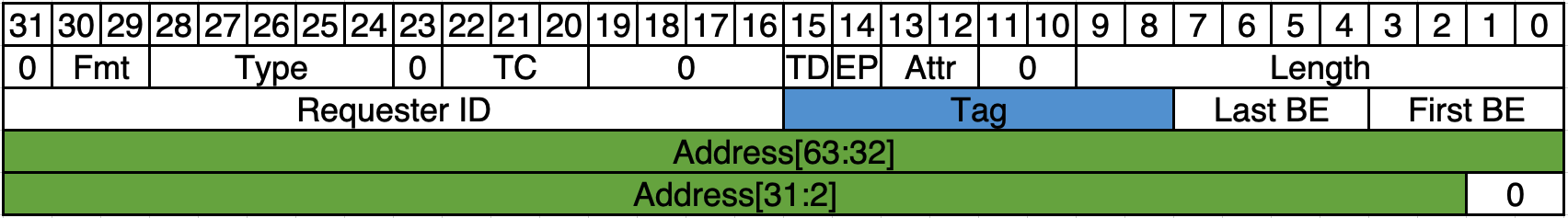}
\vspace{-1mm}
\caption{\textbf{Example of Memory Read Request TLP.} The Tag field can be used to identify the corresponding completion TLP. The address field is the targeted reading address.}
\label{fig:Request}
\end{figure}

\begin{figure}[t]
\centering
\includegraphics[width=\linewidth]{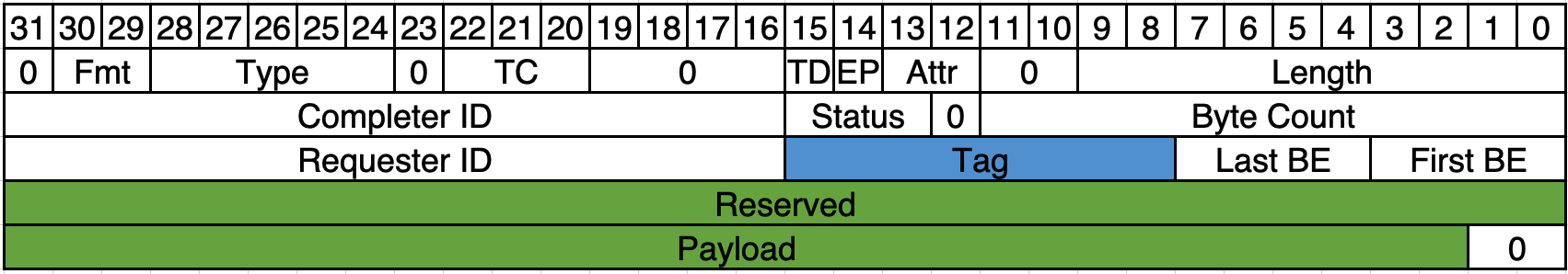}
\vspace{-1mm}
\caption{\textbf{Example of Completion TLP.} The Tag field can be used to identify the corresponding request TLP. The payload field includes the reading data from the targeted address.}
\label{fig:Completion}
\vspace{-1mm}
\end{figure}

PCIe is a high-speed motherboard interface for I/O devices, such as graphics cards, SSDs, Wi-Fi, etc. 
The communication of PCIe takes the form of packets transmitted over these dedicated lines, with flow control, error detection and re-transmissions.
The underlying communications mechanism of PCIe protocol is composed of three layers: Transaction Layer, Data Link Layer, and Physical Layer. Figure~\ref{fig:Request} and Figure~\ref{fig:Completion} show the formats of memory read request Transaction Layer Packet (TLP) and completion TLP with 64-bit addressing. The header of each TLP is four double words (DWs) long, and the maximum payload size is 128 DWs. 

When a CPU writes data into a peripheral, the chipset generates a memory write packet which consists of a 32-bit header and a payload containing the data to be written.
The packet is then transmitted to the chipset’s PCIe port. The peripheral can be connected directly to the chipset or connected to a switch network.

When a CPU reads data from a peripheral, there are two packets involved in the read operation.
One is read request TLP that is sent from CPU to the peripheral, asking the latter to perform a read operation, as shown in Figure~\ref{fig:Request}. The other one is completion TLP which comes back with data in the payload, as shown in Figure~\ref{fig:Completion}. The completion TLP and request TLP can be identified by the same Tag value.

%-------------------------------------------------------------------------------
\section{Attack Design} \label{attack}
\label{sec:figs}
%-------------------------------------------------------------------------------

\subsection{Overview}
\label{sec:overview}

\begin{figure}[t]
\centering
\includegraphics[width=\linewidth]{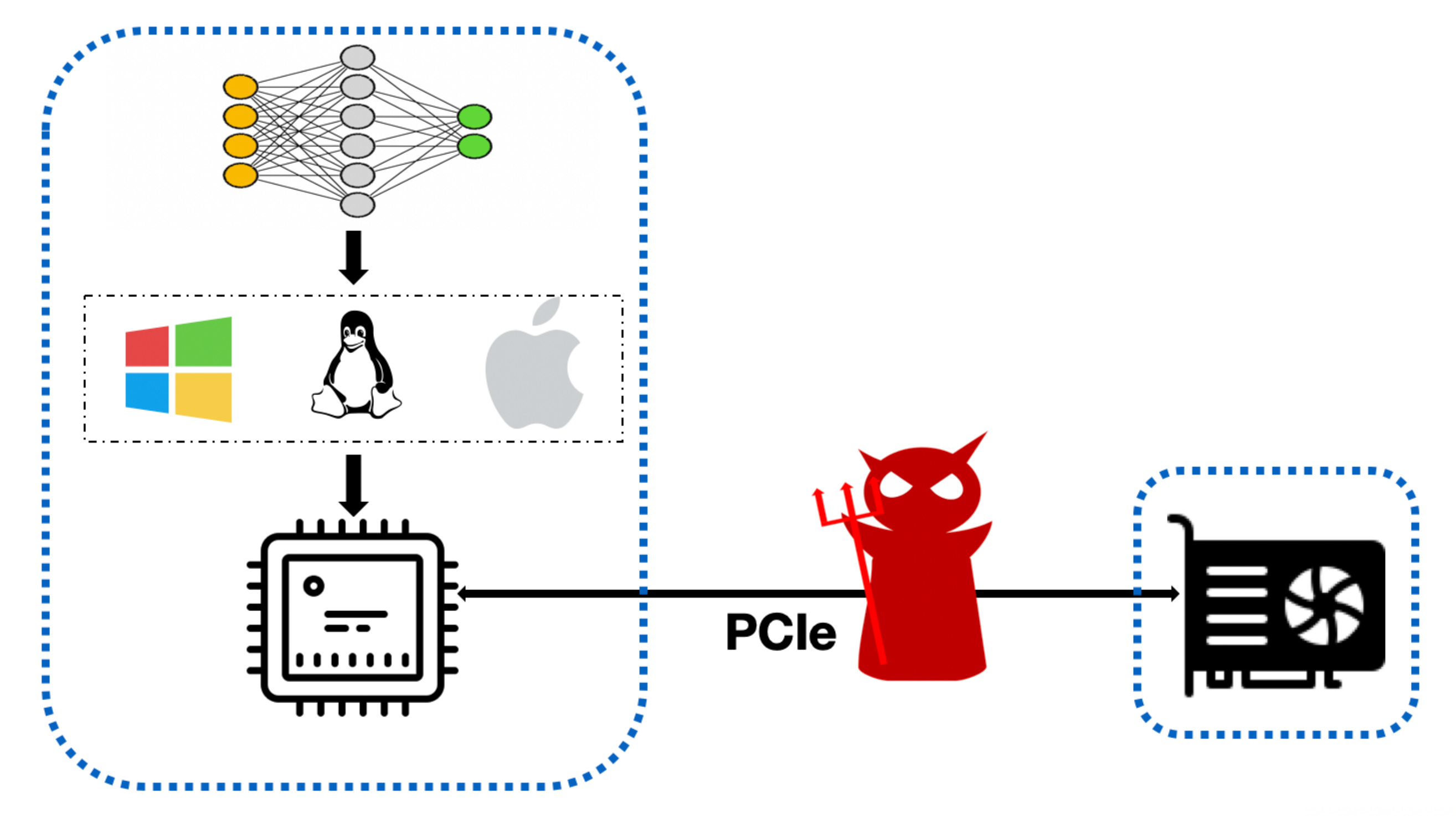}
\vspace{-1mm}
\caption{\textbf{Threat Model.} 
We consider the model privatization environment, where the host and the GPU device are well protected individually, and the PCIe bus is the new attack surface. The adversary can snoop the PCIe traffic using a bus snooping device, e.g., a PCIe protocol analyzer.}
\label{fig:threatmodel}
\end{figure}

\begin{figure*}[t]
\centering
\includegraphics[width=\linewidth]{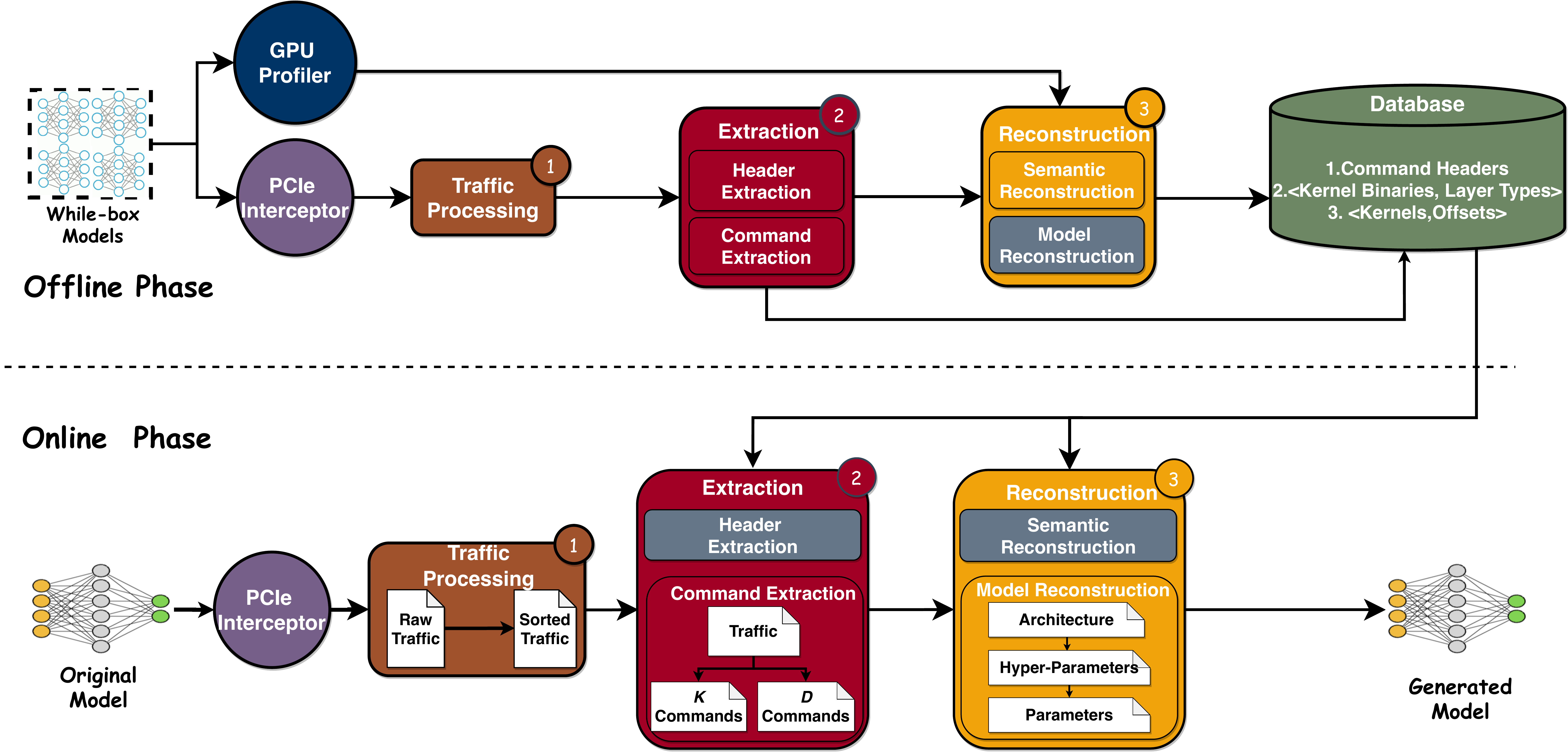}
\vspace{-2mm}
\caption{\textbf{Attack Overview.} The offline phase builds a knowledge database by identifying GPU command headers of interest, 
the mappings between GPU kernel (binaries) and DNN layer types, and the mappings between GPU kernels and offsets of hyper-parameters. The online phase is the actual deployed attack to steal the victim model during inference. Three major modules are used in both phases but with different sub-components activated (grey diagrams indicate inactivity): The traffic processing module ~\textcircled{ \scriptsize 1 } sorts out-of-order PCIe packets; 
The extraction module ~\textcircled{\scriptsize 2} extracts and filters GPU commands of interest;
The reconstruction module ~\textcircled{\scriptsize 3} fully reconstructs the semantics, architecture, hyper-parameters, and parameters.
%Note that the grey components in module ~\textcircled{\scriptsize 2} and ~\textcircled{\scriptsize 3} are not activated.
}
\label{fig:overview}
\vspace{-1mm}
\end{figure*}

\mypara{Threat Model} \label{threatmodel}
In this paper, we consider an AI model privatization deployment environment (e.g., smart IoT, surveillance devices, autonomous driving), where service providers pack their private AI models into heterogeneous CPU-GPU devices and sell them to third-party customers with subscription or perpetual licensing. The end-users are able to physically access the hardware, especially, the PCIe interface. The thread model is depicted as Figure~\ref{fig:threatmodel}, where the GPU is attached to the host via an unencrypted PCIe connection.
We assume the host and the GPU device are well protected individually, e.g., AI models are protected with existing software-hardening techniques on the host side, such as secure boot, full disk encryption, and trusted execution environment (e.g., Intel SGX~\cite{intelsgx}). 
It leaves the PCIe bus as a new attack surface for attackers. 
This assumption is reasonable in the privatization deployment environments because: (1) attackers (e.g., insiders within the third-party company) have the motivation to extract the AI model for saving the per-year license fee, and (2) attackers have physical access to the host machine, and thus they can install a PCIe bus snooping device (e.g., PCIe protocol analyzer) between the host and GPU to monitor and log the PCIe traffic.
The victim model is considered a black-box.
%, i.e., none of the victim models' information is used during our attack.
The victim can be either an existing model or a customized model. 
It can be implemented with arbitrary deep learning frameworks.

\mypara{Challenges} \label{sec:challenges}
It is challenging to fully reconstruct DNN models from PCIe traffic even if we can intercept and log all PCIe packets. We summarize the challenges as follows:
\begin{enumerate}
\item \textbf{Closed-source Code and Undocumented Data Structures}. The CUDA runtime, driver, and NVIDIA GPU hardware are all closed-source, and the critical data structures involved in data transfer and GPU kernel launch are undocumented. 
The closed-source code and per-architecture instruction set make fully disassembling impractical. Moreover, GPU kernels and commands are encoded with different instruction sets, making reverse engineering more difficult.

\item \textbf{Semantic Loss in PCIe Traffic}. Some critical semantic information of a DNN model is lost at the level of PCIe traffic. For instance, DNN layer types can not be obtained directly from PCIe traffic because it is resolved on the CPU side. The loss of critical information makes it challenging to recover the whole model fully.

\item \textbf{PCIe Packets with Numerous Noises and Chaotic Orders}. There are millions of packets generated for a single image inference, in which only 1\% to 2\% are useful for our DNN model reconstruction. The rest ``noises'' packets should be carefully eliminated.
Moreover, numerous completion packets, which indicate operation completion, often arrive out-of-order compared to DNN level semantics, due to the CUDA features that pipeline asynchronous operations.
This situation is even worse in the more advanced GPU architectures (e.g., \texttt{NVIDIA Geforce RTX 2080 Ti}) because of introducing new features to unify GPU device and host memory.
\end{enumerate}

\mypara{Attack Overview} \label{attackoverviw}
The methodology of our attack can be divided into two phases: offline phase and online phase. During the offline phase, 
we use white-box models to build a database with the identified command headers, the mappings between GPU kernel (binaries) and DNN layer types, and the mappings between GPU kernels and offsets of hyper-parameters. 
Specifically, the traffic processing module (\textcircled{\scriptsize 1} in Figure~\ref{fig:overview}) sorts the out-of-order PCIe packets intercepted by PCIe snooping device.
The extraction module (\textcircled{\scriptsize 2}) has two sub-modules: header extraction module and command extraction module. The header extraction module extracts command headers from the sorted PCIe packets (Section~\ref{sec:headerextraction}). 
The extracted command headers will be stored in the database, accelerating command extraction in the online phase. 
The command extraction module in the offline phase helps get the kernel binaries (Section~\ref{sec:commandextraction}). 
The semantic reconstruction module within the reconstruction module (\textcircled{\scriptsize 3}) takes the inputs from the command extraction module and the GPU profiler to create the mappings between the kernel (binary) and the layer type, as well as the mappings between the kernel and the offset of hyper-parameters, facilitating the module reconstruction in the online phase (Section~\ref{sec:semanticreconstruction}).

During the online phase, the original (victim) model is used for inference on a single image. The victim model is a black-box model and thoroughly different from the white-box models used in the offline phase. PCIe traffics are intercepted and sorted by the traffic processing module.
The command extraction module (\textcircled{\scriptsize 2}) extracts $K$ (kernel launch related) and $D$ (data movement related) commands as well as the GPU kernel binaries, using the header information profiled from the offline phase (Section~\ref{sec:commandextraction}).
{The entire database are feed to the model reconstruction module (\textcircled{\scriptsize 3}) to fully reconstruct architecture, hyper-parameters, and parameters (Section~\ref{sec:modelreconstruction}).}
All these steps need massive efforts of reverse engineering.

\begin{figure}[t]
\centering
\includegraphics[width=\linewidth]{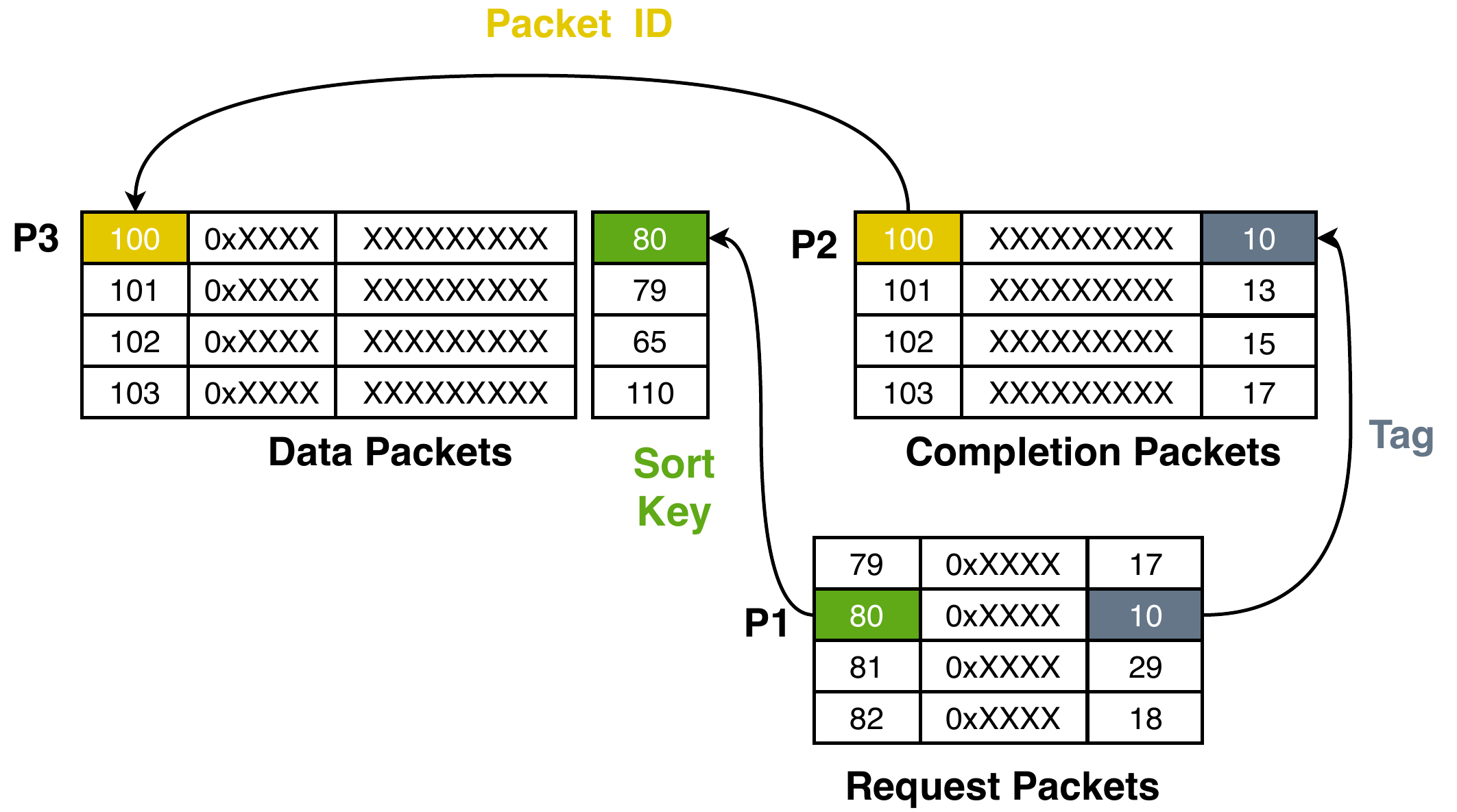}
\vspace{-2mm}
\caption{\textbf{Process of Sorting PCIe Traffic.} We sort the packets using \textit{packet ID} and \textit{tags}, instead of the capture order.}
%completion packets arrived out-of-order, and the data packets have the same order with completion packets, while the request packets can reflect the correct order. We use the packet ID in data packets get the tag from completion packets, then use tag get the correct order from request packets.
%By default, all the packets are arranged according to their packet IDs from low to high. For every packet in the upstream packets, we record it as $P_{1}$ and lookup all the downstream packets that have larger packet IDs than $P_{1}$. We stop the searching when it hits the packet that has the same tag value with $P_{1}$ and records this packet as $P_{2}$. Next, we look for the packet that has the same packet ID with $P_{2}$ in data packets and records it as $P_{3}$. Then we add the packet ID of $P_{1}$ into $P_{3}$ as a sort key. We keep doing this until every data packet has a sort key. At last, we sort all the data packets based on the sort key.}
\label{fig:example}
\vspace{-2mm}
\end{figure}

\subsection{Traffic Processing}\label{trafficprocess}
The intercepted traffic is composed of TLPs with unique packet IDs. Thanks to the oriented interception, the intercepted traffic is only formed by packets transmitted between CPU and GPU.
These packets are arranged increasing ID values in order of arrival.
Packets can be classified into upstream packets and downstream packets based on the transmitting direction. The upstream packets represent packets that are sent from GPU to CPU, e.g., GPU read request packets, or completion packets returning GPU computing results. 
The downstream packets are sent from CPU to GPU, e.g., CPU read request packets, completion packets with input data. The structures of two representative packages are shown as Figure~\ref{fig:Request} and Figure~\ref{fig:Completion}. To make things easier, we only keep the GPU read request packets in the upstream packets and the completion packets in the downstream packets.

In addition to the aforementioned two types of packets, we use another type of packet namely \textit{data packet} that is merged from request packets and completion packets according to the tag field. A data packet comprises both the request address and corresponding acquired data in a single packet. It can be concatenated to a completion packet with the same packet ID and equivalent order.

%Apart from these two types of packets, The analyzer is featured with providing another data-flow view, which is showing the packets merged from request packets and completion packets by the tag field, denoted as data packets in the rest of this paper. A data packet comprises both the request address and corresponding acquired data in a single packet, but the tag field is missed. It can be connected with a completion packet by the same packet ID and equivalent order. 

The major challenge here is that these data completion packets arrived out-of-order. The reason is that the PCIe protocol does not enforce the completion orders of multiple consecutive requests. Additionally, resultant output for a single PCIe read request may be encapsulated in multiple completion packets, making the raw packets hard to analyze directly.
To tackle the problem, we coalesce the raw packets by using \textit{merge} and \textit{sort} based on two observations: (1) every request is composed of one request packet and one (multiple) completion packet(s), where the orders of request packets can reflect the correct sequence; (2) completion packets for the same request are guaranteed to arrive in order. We elaborate the merge and sort operations as follows:

\vspace{1mm}\noindent\textbf{Merge}: For every data packet, we complement the tag field by looking up its corresponding completion packet(s). If adjacent packets have the same tag value, we merge them into a single packet by concatenating their data field.

%\vspace{1mm}\noindent\textbf{Sort}: For every packet in the upstream packets, we record it as $P_{1}$, and lookup all the downstream packets have larger packet IDs than $P_{1}$. We stop the search when hit the packet have the same tag value with $P_{1}$, record this packet as $P_{2}$. After it, we look for the packet that has the same packet ID with $P_{2}$ in data packets, record it as $P_{3}$, and add the packet ID of $P_{1}$ into $P_{3}$ as a sort key. We keep on doing this until every data packet has a sort key. At last, we sort all the data packets based on the sort key. This process is shown in Figure \ref{fig:example}.    

\vspace{1mm}\noindent\textbf{Sort}: The sort phase is illustrated as Figure~\ref{fig:example}. By default, all the packets are arranged according to their packet IDs from low to high. For request packet, we record it as $P_{1}$ and lookup all the completion packets that have larger packet IDs than $P_{1}$. We stop the searching when it hits the packet that has the same tag value as $P_{1}$ and records this packet as $P_{2}$. Next, we look for the packet that has the same packet ID with $P_{2}$ in data packets and records it as $P_{3}$. Then we add the packet ID of $P_{1}$ into $P_{3}$ as a sort key. We repeat this procedure until every data packet has a sort key. At last, we sort all the data packets by on the sort key.

\subsection{Extraction}
\label{sec:extraction}
After the preliminary processing, it's still onerous to reconstruct the model from the traffic. One of the main obstacles is that there are a large number of interference packets. For instance, making inference on a single image using MNIST model will generate 1,077,756 data packets (after filtering) on NVIDIA Geforce GT 730. 
However, only around 20,000 of them (2\%) are useful for our attack. This may be explained by the fact that CPU sends GPU numerous signals to do initialization, synchronization, etc. So it is necessary to filter out the irrelevant packets.
In order to focus on our goal of extracting DNN models, it is sufficient to pick only those \textit{$D$ commands} and \textit{$K$ commands}, representing data movement commands and kernel launch commands, respectively.

%The approach to make this goal is locating those necessary packets. As we described in section \ref{background}, GPU is controlled by CPU using commands. If we can extract the commands we need, we are able to shrink the traffic size, then easier to find the data we need. For our purpose, we only need to consider \textit{$D$ commands} and \textit{$K$ commands}, representing data movement commands and kernel launch commands, respectively.

\begin{figure}[t]
\centering
\includegraphics[width=\linewidth]{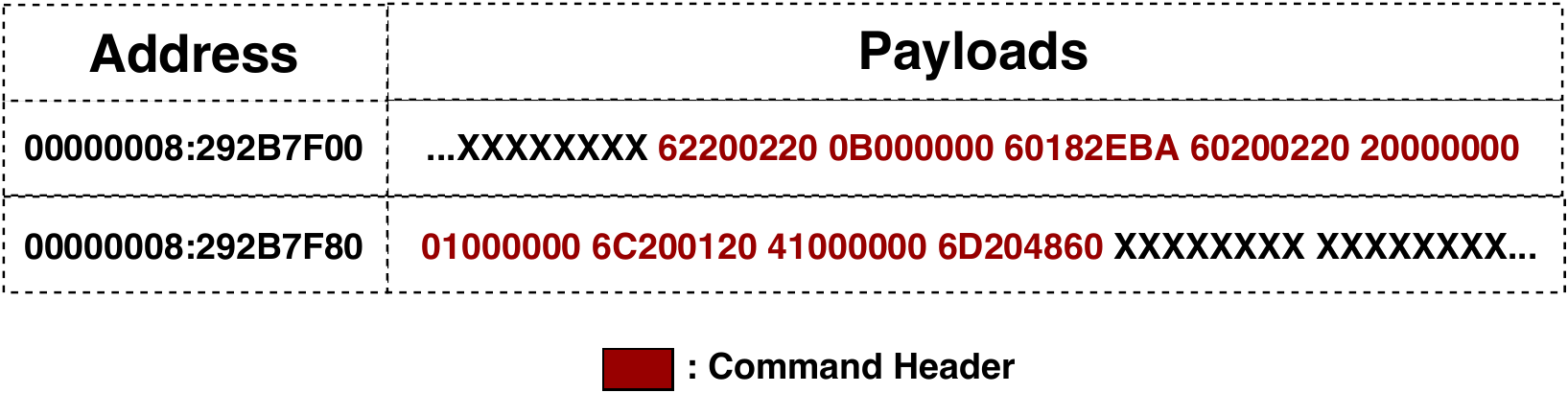}
\vspace{-2mm}
\caption{\textbf{Identified Structure of GPU Commands.} A typical GPU command consists of nine DWs. The third DW indicates the location of this command on GPU memory. The fifth DW represents the size of data field. The last DW stands for the type of this command.}
\label{fig:command}
\vspace{-1mm}
\end{figure}

\subsubsection{{Header Extraction}} 
\label{sec:headerextraction}
To extract $D$ commands and $K$ commands, we should identify the header structure of each kind of command. 
This procedure is done in our offline profiling phase.
In order to figure out the header of $D$ commands, we repeatedly move crafted data between pre-allocated GPU device memory and main memory, and use pattern match on the intercepted PCIe packets.
Similarly, we repeatedly launch multiple kernels with prepared arguments, to identify the header structure of $K$ commands.

%According to our reverse-engineeringresults, we recover the following header structure and the crit-ical fields used in our experiments.
According to our reverse-engineering results, a $K$ commands header structure is shown as the highlighted nine DWs in Figure~\ref{fig:command}, where the third, fifth, and ninth DW represents GPU memory address, data size in bytes, and command type (e.g., \textit{6D204860} is the signature indicating kernel launch on Kepler architecture), respectively. These three DW fields are most useful for our attack. The other six DWs are GPU-specific signatures whose bit-wise semantics are explained in previous reverse engineering work~\cite{koscielnicki2014envytools}.

We also did exhaustive tests to verify that the header structure is stable and valid on different GPU and machine combinations. The extracted header information are memorized in our profiling database, which can be used to accelerate analysis in the future.

%Figure~\ref{fig:command} shows an example of an extracted GPU command. The nine DWs in red font represent the command header. The third DW of this header indicates the location of this command on GPU device memory. The fifth DW represents the size of the data field in bytes, and the last DW represents the type of this command (e.g., $6D200460$ is the signature indicating kernel launch on Kepler architecture). The other six Dws are GPU-specific fingerprints.

%We find out these command headers through this method: For $D$ commands, we manually send prepared data into GPU repeatedly. By locating these prepared data, we can recognize the command header of $D$ commands. Similarly, for $K$ commands, we can launch multiple kernels with prepared arguments. By locating these arguments, we can identify the header of $K$ commands.

%\hush{Need to explain why this work, if it works for different GPUs. Better to ref Envytools rnndb.}
%According to our reverse-engineering results, we recover the following header structure and the critical fields used in our experiments.

% \begin{lstlisting}[language=C++,basicstyle=\small]
% struct Header{
%     uint32_t Fixed1
%     uint32_t Fixed2
%     uint32_t Address
%     uint32_t Fixed4
%     uint32_t Size
%     uint32_t Fixed6
%     uint32_t Fixed7
%     uint32_t Fixed8
%     uint32_t Type
% }
% \end{lstlisting}

\subsubsection{Command Extraction} 
\label{sec:commandextraction}
%Figure~\ref{fig:command} shows an example of an extracted GPU command. The nine DWs in red font represent the command header. The third DW of this header indicates the location of this command on GPU device memory. The fifth DW represents the size of the data field in bytes, and the last DW represents the type of this command (e.g., $6D200460$ is the signature indicating kernel launch on Kepler architecture). The other six Dws are GPU-specific fingerprints.

\begin{figure}[t]
\centering
\includegraphics[width=\linewidth]{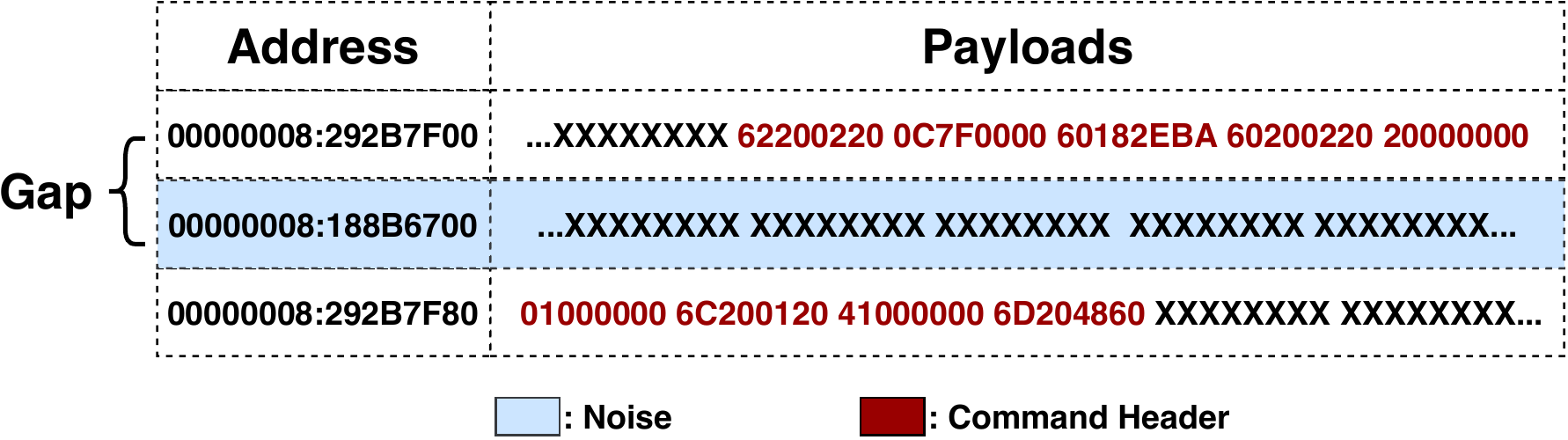}
\vspace{-2mm}
\caption{\textbf{Example of Command With Noise Packets.} The noise packet is not consecutive with the previous packet in terms of address.}
\label{fig:noise}
\vspace{-1mm}
\end{figure}

Raw extracted commands are not ready to use because of tremendous noises.
Noise can be classified into two classes: \texttt{external noise} and \texttt{internal noise}. External noise refers to those packets not belong to the current command. They can both be the packets of other commands or meaningless packets. External noise could appear frequently because a command with a large data field may require thousands of packets to transmit. Since a command header could be sent via two packets, the noise packet may also appear within the command header. As Figure~\ref{fig:noise} shown, a command header is split into two parts. They are transmitted via two packets, with a noise packet in between. Internal noise indicates a specific DW inside each packet. We have observed all internal noise and summarized the pattern of it. Thus internal noise can be easily filtered out while extracting the payloads.
%\hush{??some sentences are missing, how to address internal noise??}

\begin{figure}[t]
\centering
\includegraphics[width=\linewidth]{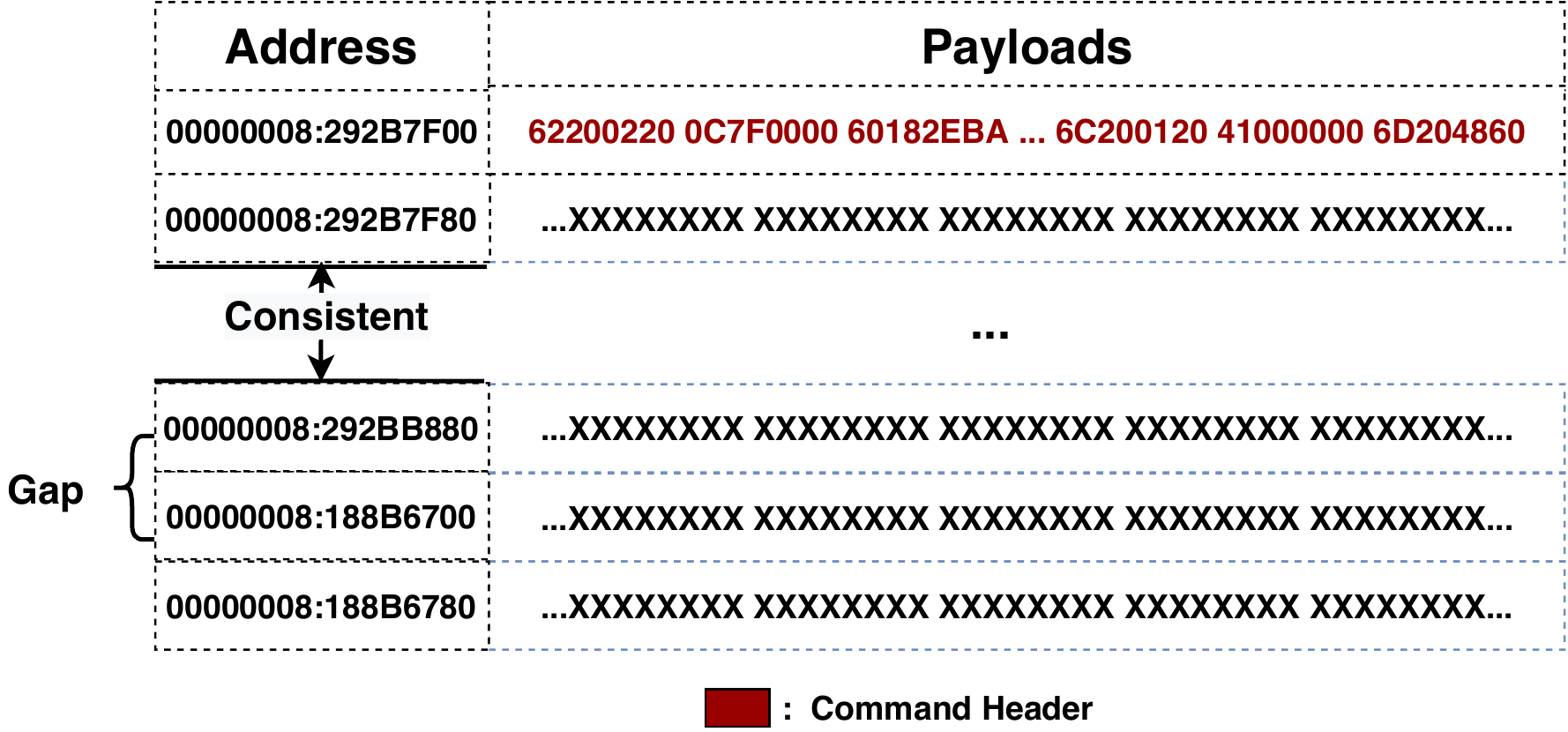}
\vspace{-2mm}
\caption{\textbf{Example of Command With Large Data Field.} When a command has a large size data field, it occupies more than one continuous memory space. In this case, the address gap also exists.}
\label{fig:Largedata}
\vspace{-1mm}
\end{figure}

An intuitive solution to address the noise issue is to check the address continuity, based on the fact that the transmitted data is usually consecutive in memory space. If a packet's memory address is not consecutive with its predecessor, it is highly likely that this packet does not belong to the current command. 
However, this is not always the case especially when the continuous memory space is insufficient. Since the addresses in packets are physical addresses, virtually contiguous address space used by CUDA programs may be split into multiple physical memory chunks.
Figure~\ref{fig:Largedata} shows an example that the addresses of two adjunct packets belong to the same command are nonconsecutive in physical address.
Therefore, it is insufficient to merely check the address continuity.
To solve this problem, we introduce a heuristic threshold \texttt{MAX\_SCAN\_DISTANCE}. When a packet encounters an address gap, we scan for the next consecutive packet within \texttt{MAX\_SCAN\_DISTANCE}. If there exists a packet that has a consecutive address with the previous address gap, we consider this packet to be the adjacent packet of the gap and discard the previously scanned packets. Otherwise, we include the gap packet into the payloads. We continue this process until the number of payloads bytes in extracted packets matches the size indicated in the command header.

\subsection{Reconstruction} 
\label{sec:reconstruction}

\subsubsection{Semantic Reconstruction} 
\label{sec:semanticreconstruction}
Semantic reconstruction is a part of the offline profiling phase to build the knowledge database. We use known DNN models as ground truth and utilize NVIDIA's profiling tools (i.e., nvprof~\cite{nvprof}) to bridge the semantic gap between PCIe packets and high-level DNN workflow by: (1) associating kernels with DNN layers; (2) profiling the layout of the arguments of certain GPU kernels. 

%The implementation of this module is challenging because some semantic information is resolved on the CPU side (e.g., layer types), which leads to the information lost in PCIe traffic. 

We assume every computational layer (e.g., convolution layer, normalization layer, rectified linear unit layer) of DNN models is computed on the GPU, because layers that are computed by CPU would not send command through PCIe.
This assumption is reasonable because the highly muti-threaded architecture of GPU is designed to accelerate matrix computation in DNN layers. Moreover, if some of intermediate layers are ported to CPU, the data movement is expensive.
Base on this assumption, it is safe to say each layer is associated with one or more GPU kernels. Different types of layers use different GPU kernels, thus we can infer the layers types by identifying their GPU kernels. Additionally, people prefer to use highly optimized standard libraries provided by GPU hardware vendors (e.g., NVIDIA's CUDNN library), so the kernel binaries are relatively stable. For example, convolution layers call $convolve\_sgemm()$ kernels whose binaries are embedded in \texttt{nv\_fatbin} section of libcudnn.so.

We have the following two observations based on our preliminary experiments:

\vspace{1mm}\noindent\textbf{Observation 1}: Each kernel is loaded onto GPU using a $D$ command, and its data field is kernel binaries.

\vspace{1mm}\noindent\textbf{Observation 2}: Each $K$ command includes an address referring to the kernel binary to be launched.

\begin{figure}[t]
\centering
\includegraphics[width=\linewidth]{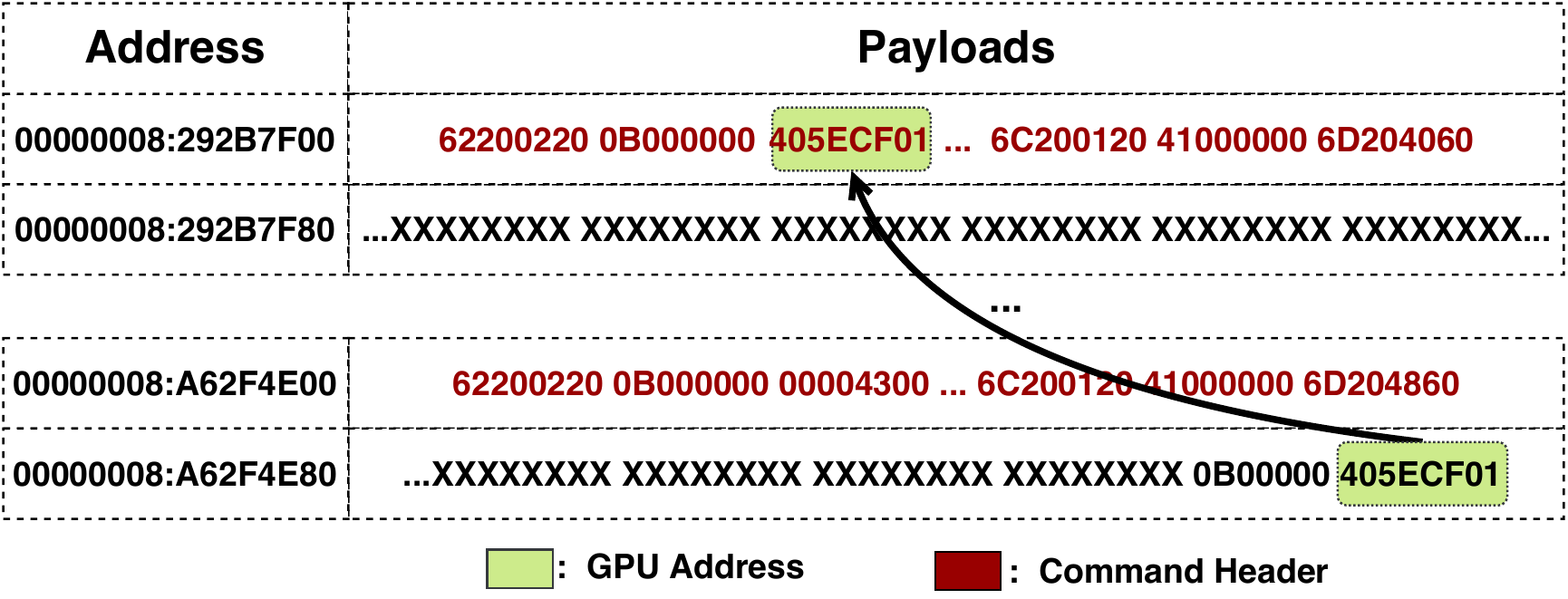}
\vspace{-2mm}
\caption{\textbf{Process of Locating Kernel Binaries.} The first command in the figure is a $D$ command that loads a kernel binary onto GPU. The second command is a $K$ command to launch the loaded kernel. These two commands are associated by the same GPU address where the kernel binary is loaded.}
\label{fig:kernellaunch}
\vspace{-1mm}
\end{figure}

Based on the two observations, we can extract all involved kernel binary by iterating $K$ commands.
Figure~\ref{fig:kernellaunch} illustrates how we use a $K$ command to locate the GPU kernel binary. 
Particularly, the kernel binary is first loaded onto GPU memory and stored at $405ECF01$ using a $D$ command, and then launched by a  $K$ command. Our method works in reverse order: we first retrieve the $K$ command's data field with a fixed offset to locate the address referring to the kernel binary, then we dump the corresponding $D$ command's data field to get the kernel binary.

%\hush{The next step is to determine the kernel type of these extracted kernel binaries. To make it, we use NVIDIA profiler-$nvprof$ to get kernel executing traces. The $nvprof$ enables the collection of a timeline of CUDA-related activities, like CUDA API calls, kernel execution, memory transfers, etc. As long as we get the kernel executing traces, we can match them with the kernel binaries by their appeared orders. At the same time, by executing different layers separately, we can get the corresponding kernels for each layer execution.}

After iterating all involved $K$ command in PCIe traffic, we have a sequence of kernel binaries in launch order. By aligning with the CUDA trace collected by nvprof, we can figure out the mappings between each kernel binary and its corresponding layer. The mappings are stored in the form of tuples in a hash table, where the key is the kernel binary and the value is layer type.

% We extract all $D$ commands and store them using a hash table, the key is the GPU address of this command, and the value is its data filed. Afterward, we scan all the packets to find $K$ commands through the command header. For every $K$ command, we check the hash table to retrieve these kernels' binaries. We continue this process until getting all executing kernels binaries. After it, we match these binaries with kernel's type (nvprof got). 
% %We first collect all existing layers' primary kernel through repeating inference procedure on different while-box training models and use $nvprof$ to obtain kernel executing trace of each layer. 
% We can determine the primary kernel of all layers in this way. At last, we store the <Kernel Binary, Layer Type> pairs into the database. Thus we know the layers.

% \Yuankun{Some kernels may not be used to identify a layer, but they can be used to identify a hyper-parameter, like activation function. Those <Kernel binary, Hyper-parameter> information will also be recovered and stored in the database.
% Apart from those pairs, the semantic reconstruction model is also responsible for identifying fixed offsets in $K$ commands, e.g., the offset of kernel binary address, the offset of hyper-parameters. These offsets can be determined by observing the data field of K commands while changing the variables for examine.}
% \hush{1. profile hyperparam offsets on differnet GPUs.}

Another semantic we need to reconstruct is the relationship between kernel binaries and their arguments layout. We only focus on the kernels that involves potential hyper-parameters. Since hyper-parameters are not parts of the trained model, they are only used in certain kernels as arguments. 
By figuring out the locations of hyper-parameters in $K$ commands, we can extract all involved hyper-parameters. We achieve this by profiling known DNNs, looping over the data field of certain kernels' $K$ commands to find the expected hyper-parameters. The <Kernels, Offsets of Hyper-parameter> pairs are recovered and stored in the knowledge database.

%Semantic reconstruction is also responsible for acquiring hyper-parameter offsets and <kernel binary, hyper-parameter> pairs.
%Some of the hyper-parameters are directly used by kernels, which implies they are sent to the GPU through $K$ commands. 
%Based on this, we can analyze the data field of the $K$ commands and find the offset of these hyper-parameters, through changing the layer's hyper-parameter and observe the changes in the data part of $K$ commands.
%These parameters may not appear in the same kernel. So we need to get all the relevant kernels of the target layer to get all the hyper-parameters. 
%The offset of each hyper-parameter may vary for different underlying GPUs. So it's necessary to re-calculate the offsets when performing an attack on a new platform. These offset information will be stored into database to facilitate the online phase.
%will change along with the change of underlying GPUs. 
%Apart from these directly obtained hyper-parameters, others could only be inferred from existing kernel information, like activation functions, use\_bias. The kernel binary -> hyper-parameter mapping information will also be stored in the database.}

\subsubsection{Model Reconstruction}
\label{sec:modelreconstruction}

\mypara{{Extract Model Architecture}}
In the online phase, after intercepting all PCIe traffic, we are able to obtain all needed $K$ and $D$ commands. The key idea of reconstructing DNN architecture is to build data flow graph where each data movement indicates an edge and every kernel launch represents a vertex. 

Every kernel takes at least one address as its input and write its output to one or more addresses.
By knowing the semantics of this kernel in profiling phase, in the form of $K$ command, we are able to figure out which offset(s) indicate input(s) and output(s). We build the data flow graph majorly by treating the input addresses as flow-from and the output addresses as flow-to. All kernels are then associated with these data addresses. We note that in the data flow graph one kernel's output address does not necessarily exactly match its successor's input address.
Because these two addresses can be within the same data block or data is copied from one address to the other, which can be determined by iterating $D$ commands.
Once the data flow graph is reconstructed, we can substitute every kernel vertices with their corresponding DNN layers by querying the mappings in the knowledge database.

\mypara{Extract Hyper-parameters} 
\label{extracthyperparameters}
The next step is to extract hyper-parameters that are used during inference, e.g., strides, kernel size. Hyper-parameters that are used to control training phase can not be captured by our inference-time attack, e.g., learning rates, batch size.
These hyper-parameters are obtained by two means. One is obtained from kernel arguments (e.g. strides) by retrieving the data fields of certain kernel launch $K$ commands, whose offsets are profiled in the semantic reconstruction step.
Another kind of hyper-parameters are determined by the existence of relevant kernels. For example, if there is a \textit{BiasNCHWKernel} kernel launch, then the boolean type hyper-parameter \texttt{use\_bias} is determined to be \texttt{true}.

% For the first kind of hyper-parameters, the offline phase will determine the fixed offset, which indicate the location of the hyper-parameter in the data field of $K$ commands, then store them into database. These offsets can be directly used to obtain their corresponding hyper-parameters during the online phase. 
% For the Second type, they can be determined by the relevant kernels. e.g. activation function can be determined by the following kernel type. The kernel binary->hyper-parameter information can also be obtained from database.

%These hyper-parameters are very similar to a single layer: they are also be resolved on the CPU side. Thus we can determine the value of these hyper-parameters according to whether the corresponding kernel exists.
%\hush{active function, follows a relu, it's a relu. use\_bias bool, following bias kernel indicate true.}

\begin{figure}[t]
\centering
\includegraphics[width=\linewidth]{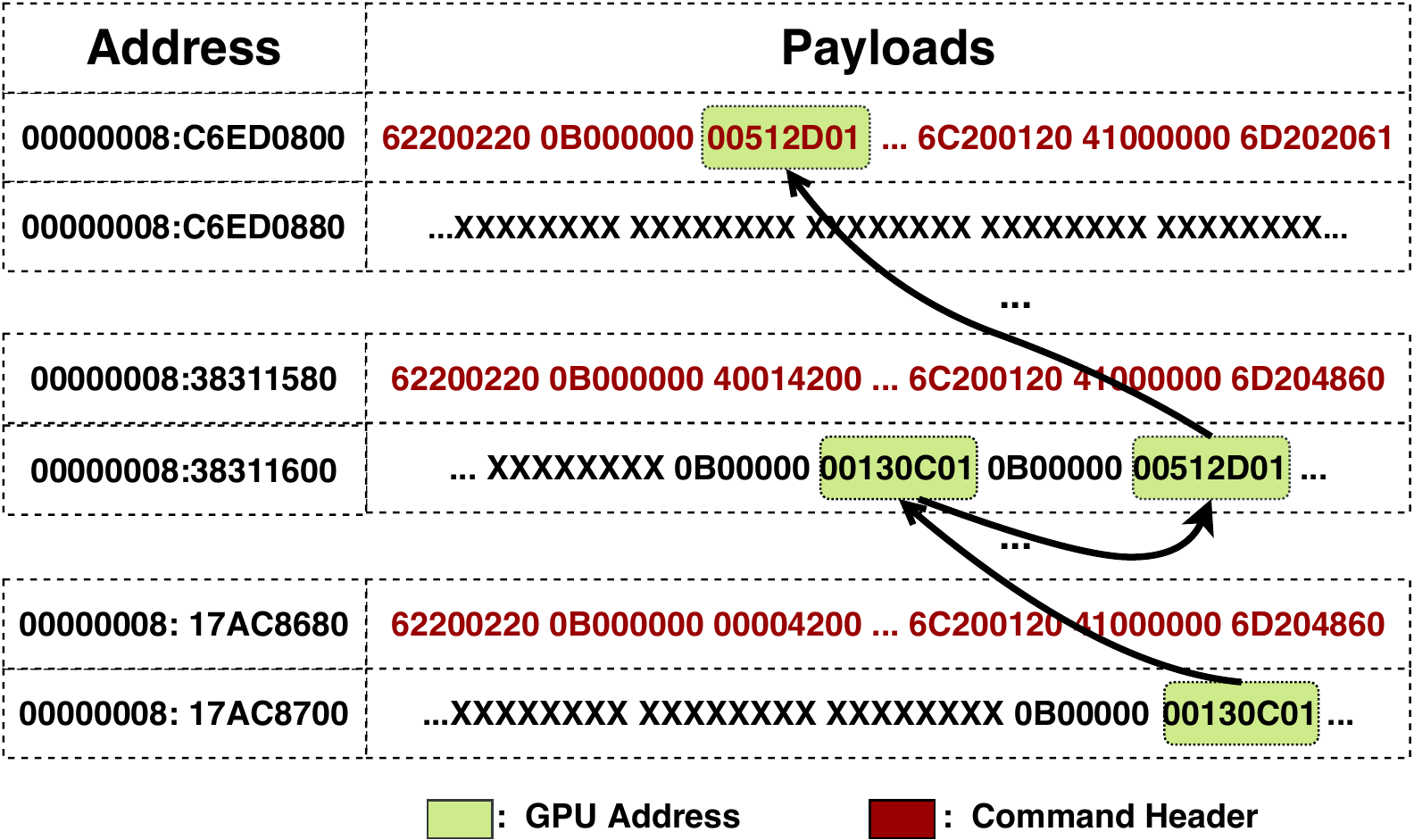}
\vspace{-2mm}
\caption{\textbf{Process of Locating Parameters.} The first command is a $D$ command of loading parameters onto GPU. The second command is a $K_{D2D}$ command which copies parameters to a new location. The third $K$ command launches a kernel taking the address of duplicated data as the input. Our attack recovers the parameters in reverse order as depicted by the arrows.
}
\label{fig:weights}
\vspace{-2mm}
\end{figure}

\mypara{Extract DNN Parameters}
In this step, we aim to obtain all the parameters of each layer. The parameter here includes both weights and bias. Intuitively, parameters are easier to obtain compared to architecture, because they are statically passed to the layer-specific APIs and propagated to the PCIe traffic in plain value. 
However, the implementations of different DNNs on different DNN frameworks vary a lot, some of them raise challenges for our attack, including duplicated parameters, asynchronous data movement, and GPU address re-use. 

The difficulty is how to locate these parameters since $D$ commands are not only used to transmit parameters but also transmit input and a lot of other data.
In our preliminary experiments, we observe that a lot of $K$ commands do not use any data that are moved onto GPU by $D$ commands. Instead, they use new addresses that are generated by certain $K$ commands. By aligning with the CUDA trace, we figure out that such $K$ commands are actually performing device to device memory copy. We name these $K$ commands $K_{D2D}$. Our understanding is that, for synchronous data copy on GPU device memory, it is much more efficient using GPU kernel than involving DMA copy which is controlled by $D$ commands. We verified our thoughts by varying test platforms and using various data size. The duplicated data are the weights of DNN layers, where the original weights on GPU memory are left untouched to avoid being polluted in inference. here comes our third observation:

\vspace{1mm}\noindent\textbf{Observation 3}: CUDA uses $K$ commands to synchronously copy data from device to device, which are named $K_{D2D}$. DNN parameters are sent to GPU using $D$ commands and often duplicated by $K_{D2D}$ commands. The data taken part in the layer computations are the copy instead of the original one.

Figure~\ref{fig:weights} illustrates how parameters are propagated among commands. The first packet (i.e., a $D$ command) is the earliest received packet by GPU. The third DW \textit{00512D01} is the GPU memory address referring to the address that stores the weights of this parameter. The second packet is a $K_{D2D}$ command where two addresses \textit{00130C01} and \textit{00512D01} in its data field. The former address is the destination and the later is the source in device to device memory copy operation. The last packet is a $K$ command launching a kernel taking the destination address as its argument. 
We recover the parameters in reverse order: (1) we first use $K$ commands to locate the destination address; (2) then we use the $K_{D2D}$ command to find the corresponding source address; (3) finally we retrieve the data field of corresponding $D$ command to dump the weights of parameters.

%D Commands can also be classified into two classes: regular $D$ Commands and \hush{async} $D$ commands. 
%\hush{handling async commands}
We found that for extremely large parameter blocks, they are usually not transmitted using regular $D$ commands. Instead, they are transferred using a new type of data movement command with different header structures. By aligning with the CUDA trace, we figure out that these commands are doing asynchronous data transfer. We name it $D_{asyn}$ command. This makes sense because the DNN framework prefers to hide the latency of large data transfer by taking it off the critical path. 
%This new kind of command will bring \hush{three??} problems toughen our reconstruction. 
New challenges are brought by $D_{asyn}$ command.
Firstly, the data size is missed in the $D_{asyn}$ command header. 
Secondly, command header and command data are located in separate packets with in-consecutive address.
%We can also get the address from it, but we can't get data size, which means we can't determine the end index of this command. Another difficulty is that we found this kind of commands can be sent simultaneously with other commands, which causes distinct parameters mixed together.\hush{yuankun rewrite}

%We designed the following approaches to address these two problems.
%\hush{1. no data size, 2. mixed}:
To resolve the first problem, we calculate the total number of weights using obtained hyper-parameters. There are three types of layers that have weights: convolution layer, dense layer, and normalization layer. 
The total number of weights and bias of convolution layers can be calculated by the following equations: 
\setlength{\belowdisplayskip}{0pt} \setlength{\belowdisplayshortskip}{0pt}
\begin{equation}\label{weights1}
    \#\ Weights_{conv} = m_w * m_h * c_{in} * c_{out}
\end{equation}
\begin{equation}\label{bias1}
    \#\ Bias_{conv} = c_{out}
\end{equation}
In Equation~\ref{weights1}, $m_w$, $m_h$ are shorted for \texttt{mask width} and \texttt{mask height}, where mask is also known as image processing kernel in convolution layers. $c_{in}$ and $c_{out}$ represent the number of input and output, which are indicated by the last arguments of input and output. $c_{out}$ is also known as $filters$.
For dense layer, the number of weights and bias can be calculated by: 
\begin{equation}\label{weights2}
    \#\ Weights_{dense} = c_{out} * c_{in}
\end{equation}
\begin{equation}\label{bias2}
    \#\ Bias_{dense} = c_{out}
\end{equation}
In Equation~\ref{weights2} and Equation~\ref{bias2}, the $c_{in}$ and $c_{out}$ represent the input shape and output shape respectively. The number of $bias$ is equal to the number of output.
%as shown in Equation.~\ref{bias1} and Equation.~\ref{bias2}.
In normalization layer, the number of weights and bias can be directly obtained from kernels' arguments without any calculation.

%To resolve the second problem, we have to make sure every packet is read only once in the correct order.  So we first sort all the layers containing parameters according to the appear order of corresponding $K$ command, then get the data of these parameters in order. While extracting data, we will mark every packet as read after extracting the payload of each packet, and keep reading until we get the same amount of weights as calculated.\hush{yuankun rewrite}

%\Yuankun{The second problem will make finding the data field of $D_{asyn}$ command difficult. For regular $D$ commands and $K$ commands, the header and data are located in the same packet, or located in two packets with consecutive addresses. We can easily find the data field of these commands. But for $D_{asyn}$, there could be a number of packets from other commands exist between the header and first packet of command data, and most of those packets are from the previous $D_{asyn}$ command. The intuition to solve this issue is, although the $D_{asyn}$ command header can mix with another $D_{asyn}$ command, the data of a $D_{asyn}$ command can be transferred only if the previous $D_{asyn}$ command completed.  In order to skip these packets, we have to make sure every packet is read only once. We first mark every packet of extracted commands as read. Then, we extract $D_{asyn}$ commands by their headers' appeared orders, and mark those packets as read. Thus, when we encounter a $D_asyn$ header, we can regard the first unread packet as the start of its data field.}

The second challenge caused by $D_{asyn}$ makes locating the data field of $D_{asyn}$ command difficult. In regular $D$ commands and $K$ commands, the header and the first piece of data are within the same packet, or located in two packets with consecutive addresses, which is easy to locate data fields.
But in $D_{asyn}$, its command header and data field can be interleaved by packets from other commands.
We resolve this issues by iterating all commands, filtering out all regular commands and noises from the beginning. Then only the $D_{asyn}$ commands are left. 
According to the fact that packets within the same command are contiguous in address, now we can easily assemble the header and the corresponding data field of every $D_{asyn}$ in order.

%\hush{??The intuition to solve this issue is, although the $D_{asyn}$ command header can mix with another $D_{asyn}$ command, the data of a $D_{asyn}$ command can be transferred only if the previous $D_{asyn}$ command completed. In order to skip these packets, we have to make sure every packet is read only once. We first mark every packet of extracted commands as read. Then, we extract $D_{asyn}$ commands by their headers' appeared orders, and mark those packets as read. Thus, when we encounter a $D_{asyn}$ header, we can regard the first unread packet as the start of its data field.}

When a large amount of data is used by the GPU, like VGG and ResNet, address re-use will occur.
That is, the data associated with the GPU address can be overwritten, and the subsequent multiple $K$ commands using the same address can refer to different data.
For example, we consider a command sequence \textcircled{1} $D_1(src)  \rightarrow \textcircled{2} K_{D2D_{1}}(src, dst_1) \rightarrow \textcircled{3} D_2(src) \rightarrow \textcircled{4} K_{D2D_{2}}(src, dst_2) \rightarrow \textcircled{5} K_1(dst_1) \rightarrow \textcircled{6} K_2(dst_2)$, where $D$ indicates $D$ command, $K_{D2D}$ indicates data copy on device, and $K$ represents $K$ command. 
In this example, data in $src$ is copied out by $K_{D2D_{1}}$ and then overwritten by $D_2$, two $K$ commands utilize data but referring to the same source address $src$.
To resolve this problem, we introduce \texttt{data life range} to represent the valid period of each data. 
The life range begins when it is written by a $D$ command and ends when it is consumed by a $K_{D2D}$ command.
Take the command sequence as the example, the life range of $dst_1$ is \textcircled{1} - \textcircled{2}, and the life rang of $dst_2$ is \textcircled{3} - \textcircled{4}.
Our strategy is to track back from every $K$ command to extract its corresponding parameters within its life range. So in the example, we extract $K_1$'s parameters in the order of \textcircled{5} \textcircled{2} \textcircled{1}.

%That is, multiple D commands use the same source address to transfer data, which is then sent to different target addresses by multiple D2D commands. For K Command, the corresponding D2D Command must be the last D2D Command to transfer data to this target address, so we can easily find the D2D Command and the source address.  It is a little tricky to find the D Command from the source address. The reason for this is that there will be redundant D commands, i.e. not all D commands will use D2D commands for transmission. Here we observe that D2D Command also contains information about the size of the data it transfers. In addition, we need to use the timing information of D Command and D2D Command, i.e. the transmission order of D Command is the same as the transmission order of D2D Command. So we filter the D Command by the size of the data first, so that it matches the number of D2D Command. Then match them in the order in which they appear.
%\hush{k command to find address, find D2D commands that refer to this address. Finally find the initial D command, then get the data field.}

\begin{figure*}[t]
\centering
\includegraphics[scale=0.8]{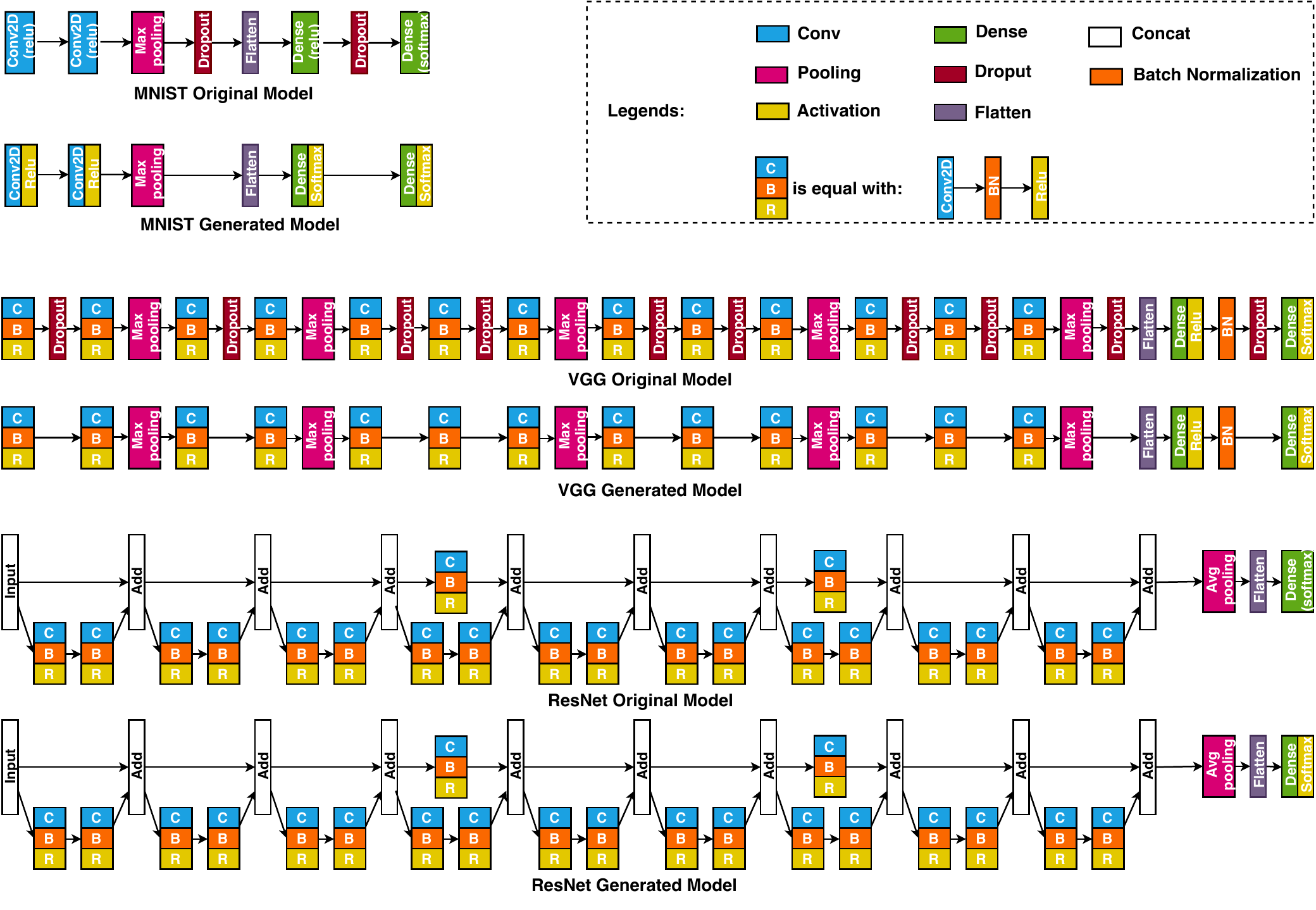}
\vspace{-1mm}
\caption{\textbf{Architecture Comparison.} This figure shows the architecture differences between original models and the reconstructed models. The CBR block represents the sequentially connected C (Convolution), B (Batch Normalization), R (Relu) layers. There are two major differences between the original models and the corresponding reconstructed models: (1) The reconstructed models do not have dropout layers (as shown in MNIST and VGG); (2) The reconstructed models treat every activation function as a single layer (as shown in MNIST and ResNet).
}
\label{fig:vggresnet_eval}
\end{figure*}

\begin{table}[t]
\caption{\textbf{Victim Models.} This table displays the detail information of all three victim models, including number of layers, number of parameters, training datasets and input shape.}
\label{tbl:vitcimmodels}
\vspace{2mm}
\resizebox{\linewidth}{!}{
\begin{tabular}{|c|c|c|c|}
\hline
                 & MNIST     & VGG16        & ResNet-20    \\ \hline
Number of  Layers    & 8     & 60    & 72   \\ \hline
Number of Parameters & 544,522   & 15,001,418 & 274,442   \\ \hline
Datasets          & mnist     & cifar10    & cifar10   \\ \hline
Input Shape      & (28,28,1) & (32,32,3)  & (32,32,3) \\ \hline 

\end{tabular}
}
\end{table}

\begin{table}[t]
                       \caption{\textbf{Related Kernels of Each Layer.} This table lists the related kernels of each layer. If there are multiple related kernels of that layer, the primary kernels are highlighted in bold. The last row indicates some kernels not belong to any layer, but are useful and need to be recorded.}
\label{tbl:relatedkernels}
\vspace{2mm}
\resizebox{\linewidth}{!}{
%\small{
\def\arraystretch{1.2}
\begin{tabular}{|c|c|c|}
\hline
\textbf{Layer}           & \textbf{Related Kernels}                        & \textbf{No.} \\ \hline
\multirow{5}{*}{Conv 2D} & \textbf{ShuffleInTensor3Simple}               & \circled{1}            \\ \cline{2-3} 
                         & cudnn::detail::implicit\_convolve\_sgemm      & \circled{2}          \\ \cline{2-3} 
                         & \textbf{SwapDimension0And2InTensor3Simple}    & \circled{3}            \\ \cline{2-3} 
                         & cudnn::winograd::generateWinogradTilesKernel  & \circled{4}            \\ \cline{2-3} 
                         & cudnn::winograd::winograd3x3Kernel            & \circled{5}            \\ \hline
BN                       & cudnn::detail::bn\_fw\_inf\_1C11\_kernel\_new & \circled{6}            \\ \hline
\multirow{2}{*}{Dense}   & \textbf{gemv2N\_kernel\_val}                  & \circled{7}            \\ \cline{2-3} 
                         & \textbf{gemvNSP\_kernel\_val}                 & \circled{8}            \\ \hline
Flatten                  & BlockReduceKernel                             & \circled{9}            \\ \hline
MaxPool                  & cudnn::detail::pooling\_fw\_4d\_kernel        & \circled{10}           \\ \hline
AvgPool                  & Eigen::internal::AvgPoolMeanReducer           & \circled{11}           \\ \hline
ZeroPad                  & PadInputCustomKernelNHWC                      & \circled{12}           \\ \hline
Add                      & Eigen::internal::scalar\_sum\_op              & \circled{13}           \\ \hline
Relu                     & Eigen::internal::scalar\_max\_op              & \circled{14}           \\ \hline
Softmax                  & softmax\_op\_gpu\_cu\_compute\_70             & \circled{15}           \\ \hline
\multirow{3}{*}{Others}  & SwapDimension1And2InTensor3UsingTiles         & \circled{16}           \\ \cline{2-3} 
                         & BiasNCHWKernel                                & \circled{17}           \\ \cline{2-3} 
                         & BiasNHWCKernel                                & \circled{18}          \\ \hline
\end{tabular}
%}
}
\end{table}

\begin{table}[t]
\caption{\textbf{Offset of Hyper-Parameters.}  This table shows all hyper-parameters offsets in their located kernel. The offset is defined as the distance between the first word and the target hyper-parameter in the data field of a $K$ command. The weights row and bias row indicate the offset of weights address and bias address respectively.}
\label{tbl:hyperparams}
\vspace{2mm}
\resizebox{\linewidth}{!}{
\def\arraystretch{1.1}
\begin{tabular}{|c|c|c|c|c|}
\hline
\textbf{Hyper-Parameters} & \textbf{Kernel}                         & \textbf{GT 730} & \textbf{1080 Ti} & \textbf{2080 Ti} \\ \hline
\multicolumn{5}{|c|}{\textbf{Convolution Layer}}                                                                                    \\ \hline
Kernel Size               & \circled{2}       & (102,103)       & (99,100)         & (96,97)          \\ \hline
Strides                   & \circled{2}       & (126,127)       & (123,124)        & (120,121)        \\ \hline
Filters                   & \circled{2}       & 101             & 98               & 95               \\ \hline
Weights                   & \circled{1}       & 83              & 80               & 80               \\ \hline
Bias                      & \circled{17}      & 85              & 82               & 82               \\ \hline
\multicolumn{5}{|c|}{\textbf{Batch Normalization Layer}}                                                                            \\ \hline
Weights1                  & \circled{6}   & 159             & 156              & 156              \\ \hline
Weights2                  & \circled{6}   & 161             & 158              & 158              \\ \hline
Weights3                  & \circled{6}   & 163             & 160              & 160              \\ \hline
Weights4                  & \circled{6}   & 165             & 162              & 162              \\ \hline
\multicolumn{5}{|c|}{\textbf{Maxpooling Layer}}                                                                                     \\ \hline
Pool Size                 & \circled{10}          & (152,153)       & (149,150)        & (146,147)        \\ \hline
Strides                   & \circled{10}          & (136,137)       & (133,134)        & (130,131)        \\ \hline
\multicolumn{5}{|c|}{\textbf{AveragePooling2D Layer}}                                                                               \\ \hline
Pool Size                 & \circled{11}          & (110,111)       & (107,108)        & (107,108)        \\ \hline
Strides                   & \circled{11}          & (114,115)       & (111,112)        & (111,112)        \\ \hline
\multicolumn{5}{|c|}{\textbf{Zeropadding Layer}}                                                                                    \\ \hline
Padding                   & \circled{12}          & (117,118)       & (114,115)        & (114,115)        \\ \hline
\multicolumn{5}{|c|}{\textbf{Dense Layer}}                                                                                          \\ \hline
Units                     & \circled{7}          & 101             & 98               & 98               \\ \hline
Weights                   & \circled{7}           & 81              & 78               & 78               \\ \hline
Bias                      & \circled{18}          & 85              & 82               & 82               \\ \hline
\end{tabular}
}

\end{table}

\section{Attack Evaluation} 
\label{evaluation}

\subsection{Experiment Setup}
\vspace{1mm}
\noindent\textbf{Hardware Platform}: We validate our attack on three GPU platforms, i.e., NVIDIA Geforce GT 730, NVIDIA Geforce GTX 1080 Ti and NVIDIA Geforce RTX 2080 Ti.
There is only one GPU attached to the motherboard via PCIe 3.0 in every individual experiment.
We adapt CUDA 10.1 as the GPU programming interface and Teledyne LeCroy Summit T3-16 PCIe Express Protocol Analyzer as our snooping device.

\vspace{1mm}
\noindent\textbf{Victim Model}: We validate our attack on three pre-trained DNN models: MNIST, VGG16, and ResNet20, which are public available~\cite{kerasapp}.
\begin{itemize}
    \setlength\itemsep{0.2mm}
    \item \textbf{MNIST} model is a sequential model, where layers are stacked and every layer takes the only output of the previous layer as the input. It is trained on the MNIST dataset and can achieve 98.25\% inference accuracy for handwritten digits.
    \item \textbf{VGG16} model is a very deep sequential model with 60 layers in total, 13 of which are convolution layers. It is trained using the cifar10 dataset and can achieve 93.59\% inference accuracy for the cifar10 test set. 
    \item \textbf{ResNet20} model is a non-sequential model, where some layers have multiple outputs and take multiple inputs from other layers. The victim ResNet model has 20 convolution layers out of 72 layers in total, which achieves 91.45\% inference accuracy for the cifar10 test set.
\end{itemize}

These pre-trained victim models are used for inference by Keras framework with Tensorflow as the backbone. In our experiments, we treat these models as black-boxes without using any layer information during attack. These publicly available models are used only for ground truth purpose. Our attack works for arbitrary proprietary models. {The attack results are not influenced by the model accuracy or architecture.}
The detailed model information including layers, shapes, and parameters are elaborated in Table~\ref{tbl:vitcimmodels}.

\begin{table}[t]
\caption{\textbf{Identity Evaluation.} This table shows the identity between the original models and the reconstructed models. All the reconstructed models have the same accuracy with the original ones, as well as similar inference time.}
\label{tbl:accuracy}
\vspace{2mm}
\resizebox{\linewidth}{!}{
\def\arraystretch{1.2}
\begin{tabular}{|c|c|c|c|c|c|}
\hline
\textbf{Metrics}                                                              & \textbf{Model} & \textbf{Original} & \multicolumn{3}{c|}{\textbf{Reconstructed}} \\ \hline
N/A                                                                           & N/A            & N/A                      & GT 730         & 1080 Ti       & 2080 Ti       \\ \hline
\multirow{3}{*}{Accuracy}                                                     & MNIST          & 98.25\%                  & 98.25\%        & 98.25\%       & 98.25\%       \\ \cline{2-6} 
                                                                              & VGG            & 93.59\%                  & 93.59\%        & 93.59\%       & 93.59\%       \\ \cline{2-6} 
                                                                              & ResNet         & 91.45\%                  & 91.45\%        & 91.45\%       & 91.45\%       \\ \hline
\multirow{3}{*}{\begin{tabular}[c]{@{}c@{}}Inference \\ Time(s)\end{tabular}} & MNIST          & 2.24                     & 2.39           & 2.52          & 2.38          \\ \cline{2-6} 
                                                                              & VGG            & 65                       & 63             & 63            & 61            \\ \cline{2-6} 
                                                                              & ResNet         & 20                       & 20             & 20            & 21            \\ \hline
\end{tabular}
}
\end{table}

\subsection{Model Architecture Evaluation}

% \usepackage{multirow}

In this section, we demonstrate the semantic equivalence between the original model and the reconstructed model. Figure~\ref{fig:vggresnet_eval} depict the architecture of original models and reconstructed models for MNIST, VGG, and ResNet, where each rectangle represents a DNN layer. 
As the figure is shown, most of the architectures of the original model and the reconstructed model are the same, except two differences. The first difference is that the reconstructed model does not have the dropout layers, e.g., the MNIST model and the VGG model. The dropout layer is used to prevent over-fitting during the training procedure. It randomly selects some neurons and drops the results. Since it is only used in the training phase and disabled during the inference, this information is not able to be captured in PCIe traffic. Attributes to the quiescence in interference, the dropout layer will not influence the result of the inference. The second difference is caused by the implementation. Some models are implemented using activation function as a hyper-parameter, like the original MNIST model, but some others regard activation function as a single layer, like the Relu function in the original VGG model. This implementation difference will also not lead to any accuracy variance. During our reconstruction, we regard all activation functions as single layers.

Table~\ref{tbl:relatedkernels} lists all the related kernels of each layer. Some kernels are primary kernels, and some kernels are used to obtain the offset of hyper-parameters. If a layer has only one related kernel, then this kernel is its primary kernel. If a layer has more than one related kernels, its primary kernels are highlighted in bold. The last row indicates some kernels not belong to any layer, but are still useful and need to be recorded. \textit{SwapDimension1And2InTensor3UsingTiles} is record in order to recover the data flow. \textit{BiasNCHWKernel} and \textit{BiasNHWCKernel} are used to determine the layer use bias or not and also used to obtain the offset of bias address.

%The last column in Table~\ref{tbl:reconstructedmodel} shows the identified primary kernels used to determine the layer type. The primary kernel can be a combination, e.g., convolution 2D layer can be identified by the appearance of the combination kernels: tensorflow::functor::ShuffleInTensor3Simple() and kernel cudnn::detail::implicit\_convolve\_sgemm (). In most cases, the layer can be determined by a single kernel, e.g., the dense layer can be identified by kernel gemv2N\_kernel\_val().

\begin{table*}[t]
\caption{\textbf{Performance Evaluation.} This table displays both runtime statistics and generation time. The runtime statistics include the number of extracted D Commands, K Commands, as well as the number of completion packets. Generation time in minutes refers to the time used to reconstruct the model. The inference time in seconds indicates the time used to test 10,000 images.}
\label{tbl:performance}
\vspace{2mm}
\resizebox{\linewidth}{!}{
\def\arraystretch{1.2}
\begin{tabular}{|c|c|c|c|c|c|c|c|c|c|}
\hline
                         & \multicolumn{3}{c|}{\textbf{MNIST}} & \multicolumn{3}{c|}{\textbf{VGG}} & \multicolumn{3}{c|}{\textbf{ResNet}} \\ \hline
Platform                 & GT 730     & 1080 Ti    & 2080 Ti   & GT 730    & 1080 Ti   & 2080 Ti   & GT 730    & 1080 Ti     & 2080 Ti    \\ \hline

\# of D Commands         & 25,680     & 28,590     & 24,342    & 27,287    & 27,677    & 24,931    & 28,433    & 28,518      & 25,577     \\ \hline
\# of K Commands         & 216        & 139        & 181       & 903       & 628       & 793       & 1011      & 886         & 988        \\ \hline
\# of Completion Packets & 1,077,756  & 2,244,115  & 2,959,613 & 4,284,946 & 2,615,895 & 3,354,411 & 975,257   & 2,052,657   & 2,717,451  \\ \hline
Generation Time (min)    & 5          & 8          & 11        & 17        & 11        & 12        & 6         & 9           & 10         \\ \hline

\end{tabular}
}
\end{table*}

\subsection{Hyper-Parameters Evaluation}

The extracted hyper-parameters are the same as those in the original model. Table~\ref{tbl:hyperparams} represents all hyper-parameters offsets in their located kernel. The offset is defined as the distance between the first word and the target hyper-parameter in the data field of a $K$ command. Meanwhile, we also record the weights and bias offset, which indicate the offset the weights address and bias address respectively. As Table~\ref{tbl:hyperparams} shown, the offset of these hyper-parameters is not fixed on distinct platforms. Some layers may also have multiple implementations, and the related kernels may change along with the implementation changes. Here we only list the most frequently used implementation and their offsets.

%For the victim model, three types of layers are included in hyper-parameters: Convolution layer, Maxpooling layer, and Dense layer. We record the offset of all hyper-parameters of these three layers. For the convolution layer, the filter size is equal to the first number of output shapes because the filter precisely indicates the number of outputs. For the Dense layer, units represent the output shape, so the value of these two parameters are the same. It can also be seen from 

%The value of $Activation \  function$ and $use\_bias$ are determined by primary kernels. For $Activation \  function$, the existing of a its primary kernel represent this $activation \ function$ is adopt. For  $use\_bias$, the existing of a its primary kernel represent $use\_bias$ is true. These primary kernels are depicted in Table~\ref{tbl:primarykernel}. A layer or hyper-parameter can have multiple primary kernels, e.g., $use\_bias$ can be determined by both tensorflow::BiasNCHWKernel () and tensorflow::BiasNHWCKernel ().

\subsection{Identity Evaluation} 
\label{sec:performance}

{Table~\ref{tbl:accuracy} evaluates the identity between the original models and reconstructed models. We evaluate the identity from two aspects, accuracy and inference time. The accuracy is measured as the average test accuracy on 10,000 test images.  The inference time in seconds indicates the total time used to test 10,000 images using this model. For MNIST, the test datasets is obtained from \textit{keras.datasets.mnist.load\_data}. For VGG and ResNet, the test datasets is obtained from \textit{keras.datasets.cifar10.load\_data}. The reconstructed models proved to be as accurate as of the victims on all platforms. The original MNIST model trained on the MNIST dataset achieve 98.25\% accuracy. The original VGG model and ResNet trained on cifar10 dataset achieve 93.59\% and 91.45\% respectively, and all reconstructed VGG models are ResNet models have the same accuracy with the original models.  As Table~\ref{tbl:accuracy} shown, each reconstructed model has a similar inference time with the original one, within a reasonable variance.}

\subsection{Reconstruction Efficiency} \label{sec:accuracy}
%\yueqiang{This section should include: 1) how efficient to reconstruct the AI model (only need 1 interference with 1 image); 2) runtime commands; and 3) table 5 info.}

%\begin{table}[t]
%\caption{\textbf{Performance Evaluation.} This table displays both generation time and inference time. Generation time in minutes refers the time used to reconstruct the model. The inference time in seconds indicates the time used to test 10000 image using this model.}
%\label{tbl:performance}
%\vspace{1mm}
%\resizebox{\linewidth}{!}{
%\begin{tabular}{|c|c|c|c|c|c|}
%\hline
%       & \multicolumn{3}{c|}{\textbf{Generation Time (min)}} & \multicolumn{2}{c|}{\textbf{Inference Time (s)}} \\ \hline
%       & GT 730          & 1080 Ti         & 2080 Ti         & Original               & reconstructed               \\ \hline
%MNIST  & 5               & 8               & 11              & 4                      & 4                       \\ \hline
%VGG    & 17              & 11              & 12              & 67                     & 67                      \\ \hline
%ResNet & 6               & 9               & 10              & 19                     & 19                      \\ \hline
%\end{tabular}
%}
%\end{table}

{Table~\ref{tbl:performance} records the runtime statistics and the model-generation time. The runtime statistics include the number of total completion packets and the number of both $D$ commands and $K$ commands. These statistics are obtained from the inference procedure on a single image. Only one image is enough to reconstruct the whole model. As the table shows, the number of $D$ commands does not have many relationships with the running models, since only a few $D$ commands are used to transfer the information of victim models. However, more complicate the victim model is, more $K$ commands will be involved. The generation time in minutes represents the total time used to reconstruct a model from the PCIe data, including Traffic Processing, Command Extraction, and Reconstruction. The generation time mainly relies on the number of completion packets. The number of completion packets is dependent on both platform and the victim model.
}

%Table~\ref{tbl:performance} record the generation time and the inference time. The reconstructed time in minutes stands for the total time used to reconstruct a model from the PCIe data, including Traffic Processing, Command Extraction, and Reconstruction. The inference time is recorded as the total time test 10,000 images using the model. As the table shown, the inference time of reconstructed models are exactly the same as the original model.

\begin{table*}[t]
\label{tbl:relatedwork}
\caption{\textbf{Related Work Comparison.} \checkmark stands for fully recover, P stands for partial recover, $\times$ means cannot recover.}
\vspace{2mm}
\resizebox{\linewidth}{!}{
\def\arraystretch{1.1}
\begin{tabular}{|c|c|c|c|c|c|}
\hline
\multirow{2}{*}{\textbf{Work}} & \multirow{2}{*}{\textbf{Information Source}} & \multirow{2}{*}{\textbf{Method}} & \multicolumn{3}{c|}{\textbf{Results}} \\ \cline{4-6}
                      &                                     &                         & \textbf{Architecture} & \textbf{Hyper-Parameters} & \textbf{Parameters}  \\ \hline
Xing Hu, et al. 2019 \cite{hu2019neural}           & Bus Access Pattern   & Predict              & P          & $\times$      &$\times$    \\ \hline
Yan, Mengjia, et al. 2018 \cite{yan2020cache}      & Cache                & Search               & $\times$     & $\checkmark$ & $\times$   \\ \hline 
Weizhe Hua, et al. 2018 \cite{hua2018reverse}      & Accelerator          & Search,Infer         & $\checkmark$ & $\times$     & P        \\ \hline
Yun Xiang et al. 2019 \cite{xiang2019open}         & Power                & Predict              & $\checkmark$ & $\checkmark$ & $\times$  \\ \hline
Vasisht Duddu et al. 2018 \cite{duddu2018stealing} & Timing               & Search               & $\checkmark$ & $\times$    & $\times$   \\ \hline
Binghui Wang et al. 2019 \cite{wang2018stealing}   & Parameters           & Infer                & $\times$     & $\checkmark$ & $\times$   \\ \hline
Seong Joon Oh et al. 2018 \cite{oh2019towards}     & Queries              & Infer                & P          & P          & $\times$   \\ \hline 
Roberts, Nicholas et al. 2018 \cite{roberts2019model} & Noise Input       & Predict,Infer        & $\times$   & $\times$   & P          \\ \hline 

\textbf{Our Work (Hermes Attack)} & \textbf{PCIe Bus} & \textbf{Infer} & \textbf{\checkmark} & \textbf{\checkmark} & \textbf{\checkmark} \\ \hline

\end{tabular} 
}

\end{table*}

\section{Discussions} \label{discussion}
The \name aims to leak the victim model through PCIe traffic with lossless inference accuracy. It means that the extracted model will have the same accuracy as the victim one, regardless of the victim model's accuracy. Meanwhile, the number of the activation functions and the model layers will not affect our attack's accuracy.
%\yueqiang{add compressed NN models here.}

\subsection{Super Large DNN Models}
The methodology of our attack is supposed to be effective for all models. However, the buffer size of the snooping device could be a potential limitation. We currently use the Teledyne LeCroy Summit T3-26 PCIe protocol analyzer as our snooping device, which is equipped with an 8GB memory buffer (4GB for each direction). Due to the buffer size limitation, we cannot intercept all the traffic if the size of a victim model is super large, i.e., VGG16 trained from ImageNet\cite{deng2009imagenet}. Although the size of this model is about 500MB, the generated downstream traffic will slightly exceed the buffer limitation due to the large amount of metadata generated by PCIe and GPU. This problem could be solved by updating the snooping device. As far as we know, some other powerful snooping devices like Teledyne LeCroy's Summit T34 PCI Express protocol analyzer\cite{SummitT34} can expand the memory buffer into 64GB. These devices would be able to intercept all the inference traffic of existing DNN models. Alternatively, we can  address this issue with an advanced algorithm. Specifically, although the intercepted model is not complete (e.g., only covering the first $n$ layers) , we can still run our existing algorithm mentioned in this above to recover the first $n$ layers of the model. In the next time, we try to intercept the AI model by skipping $k$ layers ($k \leq n$), and run the algorithm again. By repeating this step until we can recover the last layer, we then get the whole model by merging all existing recovered layers. This solution does not rely on any advanced hardware device, 
but it requires accurate model interception, and how to directly recover layers without the data of the skipped layers.

\subsection{Attack Generalization}
\label{generalization}
We have demonstrated that our attack can be applied to different GPU platforms. For different platforms (e.g., a smartphone with Neural Processing Unit (NPU)), there are several changes that should be noticed. The first change is the command header that could be different. One possible solution is to use the method we mentioned in Section~\ref{sec:headerextraction} to identify the new command header structure. The second change is the GPU instruction sets. The change of instruction sets will lead to the difference in kernel binaries. Fortunately, we can also use the method in Section~\ref{sec:semanticreconstruction} to update the database. Although there would be several changes when the platform changes, the GPU and PCIe underlying working mechanism will stay the same. Therefore, the proposed attack will not be influenced by the alternation of hardware. 
%For other similar scenarios, such as a smartphone with Neural Processing Unit (NPU).

Different from the change of GPUs, the change of the DNN framework will lead to the different implementation of each layer as well as the relationship between layer and GPU kernels. However, as long as all layers are executed on GPU, we are able to obtain the relationship between the layer and kernels, it will not affect our proposed attack.

The case that multiple tasks simultaneously run on a single GPU should also be aware of. The simultaneously running tasks share the same GPU with the victim model. In this manner, the data sent from the other tasks will make an interference on our extraction. Thanks to the fact that each process owns a GPU context and each context has at least one channel to sent commands, the different tasks can be filtered by the context information.

\subsection{Mitigation Countermeasures}
The first possible defense approach is to encrypt the PCIe traffic. It is easy to add the crypto engine on the CPU side, but it is hard for the commodity GPUs that do not have such capabilities. Thus, this method is the lack of backward compatibility. Another approach is to use data obfuscation, e.g., obfuscating the commands, model commands, and parameters. However, this method requires kernels to be extended to deobfuscate the data back or understand the obfuscated data. Besides, this method can only increase the bar but cannot prevent the Hermes attack completely.

Besides encryption and obfuscation, another mechanism is adding noise from the software aspect, e.g., sending data in one process but sending interference commands from a different process. However, this could be resolved by utilizing GPU channels, as discussed in Section~\ref{generalization}. Another alternative solution is to leverage the device driver to use dynamic command headers instead of static command ones, significantly increasing the bar of reverse engineering.

The last possible defense mechanism is to offload some tasks to the CPU. In this way, it can reduce the information obtained from the PCIe traffic. Unfortunately, it will result in significant performance loss due to the frequent data transfer between CPU and GPU and CPU's low computing power compared to GPU.

\section{Related Work} \label{relatedwork}
%The DNN security works are mainly from two aspects:

\vspace{-1mm}\noindent\textbf{{Adversarial Examples}}:\ {Adversarial examples are first pointed out by Szegedy et al. \cite{szegedy2013intriguing}, which are able to cause the network to misclassify an image.} They proposed the L-BFGS approach to generate adversarial examples by applying a certain imperceptible perturbation, which is found by maximizing the network's prediction. Afterward, there has been a lot of work concentrating on the adversarial attack, some of them is white-box attack\cite{szegedy2013intriguing,carlini2017towards,kurakin2016adversarial,biggio2013evasion}, that the attacker has some prior knowledge of the internal architecture or parameters of the victim model, some of the attacks are black-box attack\cite{papernot2016transferability, cheng2018query,papernot2017practical, chen2017zoo, cisse2017houdini, sarkar2017upset, bhagoji2017exploring}.

\vspace{1mm}\noindent\textbf{Extraction Attack}: Table~\ref{tbl:relatedwork} summarized some other DNN model extraction attacks and compared them with our work. \cite{hu2019neural} proposed an attack by hearing the memory bus and PCIe hints, built a classifier to predict the DNN model architecture, \cite{yan2020cache} introduced a cache-based side-channel attack to steal DNN architectures, \cite{hua2018reverse} performed a side-channel attack to reveal the network architecture and weights of a CNN model based on memory access patterns and the input/output of the accelerator, \cite{xiang2019open} revealed the internal network architecture and estimated the parameters by analyzing the power trace. Similarly,\cite{wei2018know} presented an attack on an FPGA-based convolutional neural network accelerator and recovered the input image from the collected power traces. \cite{duddu2018stealing} proposed an extraction attack by exploiting the side timing channels to infer the depth of the network. \cite{wang2018stealing} designed an attack on stealing the hyper-parameters of a variety of machine learning algorithms, this attack is derived by know parameters and the machine learning algorithms, and training data set.
\cite{hunt2020telekine} demonstrates an attack that predicts the image classify results by observing the GPU kernel execution time. \cite{roberts2019model} assumed the model architecture is known, and the softmax layer is accessible, then proved noise input is enough to replicate the parameters of the original model. 
\cite{shokri2017membership} designed a membership inference attack to determine the training datasets based on prediction outputs of machine learning models.
\cite{tramer2016stealing} investigated the extraction attack on various cloud-based ML model rely on the outputs returned by the ML prediction APIs. Similarly, some works generated a clone model from the query-prediction pairs of the victim model.\cite{orekondy2019knockoff,kariyappa2020maze,oh2019towards,tramer2016stealing,shokri2017membership}.

%\cite{orekondy2019knockoff} proposed Knockoff, which is trained by image queries-prediction pairs of the victim model.
%\cite{kariyappa2020maze} proposed a data-free model stealing attack using zeroth-order optimization, which makes the clone model achieve 0.91x to 0.99x accuracy.  
%\cite{oh2019towards} introduced a method for exposing the internals of black-box models by sending queries to the victim and read off outputs, and exposed internals can be exploited to strengthen adversarial examples against the model.
%Some other works similarly generated a clone model from the query-prediction pairs of the victim model.\cite{orekondy2019knockoff,kariyappa2020maze,oh2019towards,tramer2016stealing,shokri2017membership}.

\section{Conclusion} \label{conclusion}

In this paper, we identified the PCIe bus as a new attack surface to leak DNN models. 
Based on this new attack surface, we proposed a novel model-extraction attack, named \name,
which is the first attack to fully steal the whole DNN models.
We addressed the main challenges by a large number of reverse engineering and reliable semantic reconstruction, 
as well as skillful packet selection and order correction.
We implemented a prototype of the \name, and evaluated it on three real-world NVIDIA GPU platforms. 
The evaluation results indicate that our scheme could handle customized DNN models and the stolen models had the same inference accuracy as the original ones. 
We will open-source these reverse engineering results, hoping to benefit the entire community.

%-------------------------------------------------------------------------------

%\Urlmuskip=0mu plus 1mu\relax
\bibliographystyle{plain}
\bibliography{hermes}
% \onecolumngrid
%\clearpage

%\input{revisionletter.tex}
%%%%%%%%%%%%%%%%%%%%%%%%%%%%%%%%%%%%%%%%%%%%%%%%%%%%%%%%%%%%%%%%%%%%%%%%%%%%%%%%
\end{document}